\documentclass{svmult}
\usepackage{makeidx}         
\usepackage{graphicx}        
\usepackage{multicol}        
\usepackage[bottom]{footmisc}

\makeindex                   

\usepackage{amsmath}
\usepackage{amssymb}
\usepackage{rotating}
\usepackage{eucal}
\usepackage{rotating}

\newcommand{\reals}{\mathbb{R}}
\newcommand{\integers}{\mathbb{Z}}
\newcommand{\complex}{\mathbb{C}}

\newcommand{\tr}{\mathrm{Tr}}

\newcommand{\EmbSpace}{\mathrm{Emb}(\Sigma{,}M)}
\newcommand{\Emb}{\mathcal{E}}

\newcommand{\DiffF}{\mathrm{Diff_F}(\Sigma)}
\newcommand{\DiffbarF}{\mathrm{Diff_F}(\bar\Sigma)}
\newcommand{\DiffbarFconn}{\mathrm{Diff_F^0}(\bar\Sigma)}

\newcommand{\Riem}{\mathrm{Riem}(\Sigma)}
\newcommand{\Riembar}{\mathrm{Riem}(\bar\Sigma)}

\newcommand{\RP}{\mathbb{R}\mathrm{P}}

\newcommand{\G}{\mathcal{G}}
\newcommand{\Hcal}{\mathcal{H}}
\newcommand{\Dcal}{\mathcal{D}}

\begin{document}
\title*{Matter from space}
\author{Domenico Giulini}
\institute{Institute for Theoretical Physics, University of Hannover\\ 
Appelstrasse\,2, D-30167 Hannover, Germany; and\\
Center of Applied Space Technology and Microgravity (ZARM),\\
University of Bremen, Am Fallturm\,1, D-28359 Bremen, Germany.\\
\texttt{giulini@itp.uni-hannover.de}}

\maketitle

\begin{abstract}
General Relativity offers the possibility to model attributes of 
matter, like mass, momentum, angular momentum, spin, chirality etc. 
from pure space, endowed only with a single field that represents
its Riemannian geometry. I review this picture of `Geometrodynamics'
and comment on various developments after Einstein.   
\end{abstract}

\section{Introduction}
\label{sec:Introduction}
Towards the end of his famous habilitation address, delivered on 
June\,10th 1854 to the Philosophical Faculty of the University 
of G\"ottingen, Bernhard Riemann applied his mathematical ideas to 
physical space and developed the idea that it, even though of 
euclidean appearance at macroscopic scales, may well have a 
non-euclidean geometric structure in the sense of variable 
curvature if resolved below some yet unspecified microscopic scale. 
It is remarkable that in this connection he stressed that the 
measure for geometric ratios (``Massverh\"altnisse'') would already 
be encoded in the very notion of space itself if the latter were 
considered to be a discrete entity, whereas in the continuous case 
the geometry must be regarded as being a contingent structure 
that depend on ``acting forces''.%
\footnote{``Es mu\ss\ also entweder das dem Raume zugrude liegende
Wirkliche eine diskrete Mannigfaltigkeit bilden, oder der Grund der 
Ma{\ss}verh\"altnisse au{\ss}erhalb, in darauf wirkenden bindenden
Kr\"aften gesucht werden.''(\cite{Riemann:UeberDieHypothesen}, p.\,20)} 

This suggestion was seized and radicalised by William Kingdon 
Clifford, who in his paper `On the Space-Theory of Matter',
read to the Cambridge Philosophical Society on February 21st
1870, took up the tough stance that \emph{all} material 
properties and happenings may eventually be explained 
in terms of the curvature of space and its changes. In this 
seminal paper the 24-year old said 
(\cite{Clifford:MathematicalPapers}, reprinted on p.\,71 of 
\cite{Pesic:2007}): 
\begin{quote}
``I wish here to indicate a manner in which these speculations 
[Riemann's] may be applied to the investigation of physical 
phenomena. I~hold in fact:
\begin{enumerate}
\item
That small portions of space \emph{are} in fact 
of a nature analogous to little hills on a surface 
which is on the average flat; namely, that the 
ordinary laws of geometry are not valid in them.
\item
That this property of being curved or distorted
is continually being passed from one portion of space 
to another after the manner of a wave.
\item
That this variation of the curvature of space is 
what really happens in that phenomenon which we call 
the \emph{motion of matter}, whether ponderable or 
etherial.
\item
That in the physical world nothing else takes place 
but this variation, subject (possibly) to the law of 
continuity.''
\end{enumerate}
\end{quote}

\begin{figure}[ht]
\centering
\includegraphics[width=0.50\linewidth]{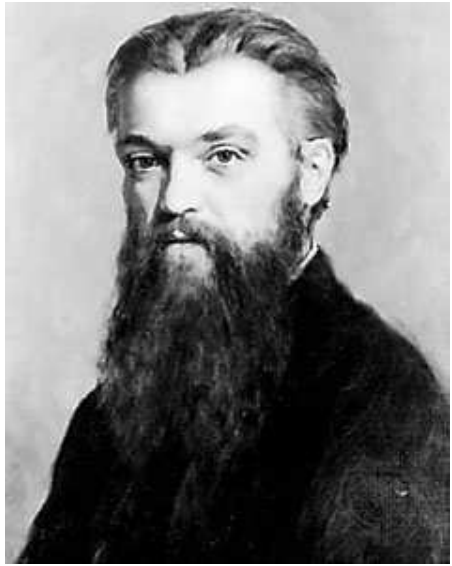}
\caption{\label{fig:Clifford}
Replica (by John Collier) of the portrait of William Kingdon 
Clifford at the London National Portrait Gallery.}
\end{figure}

In this contribution I wish to explain and comment on the 
status of this programme within General Relativity. This 
is not to suggest that present day physics offers even the 
slightest hope that this programme - understood in its radical
sense - could succeed. But certain aspects of it certainly 
are realised, sometimes in a rather surprising fashion, and 
this is what I wish to talk about here. 

\begin{figure}[ht]
\centering
\includegraphics[width=0.65\linewidth]{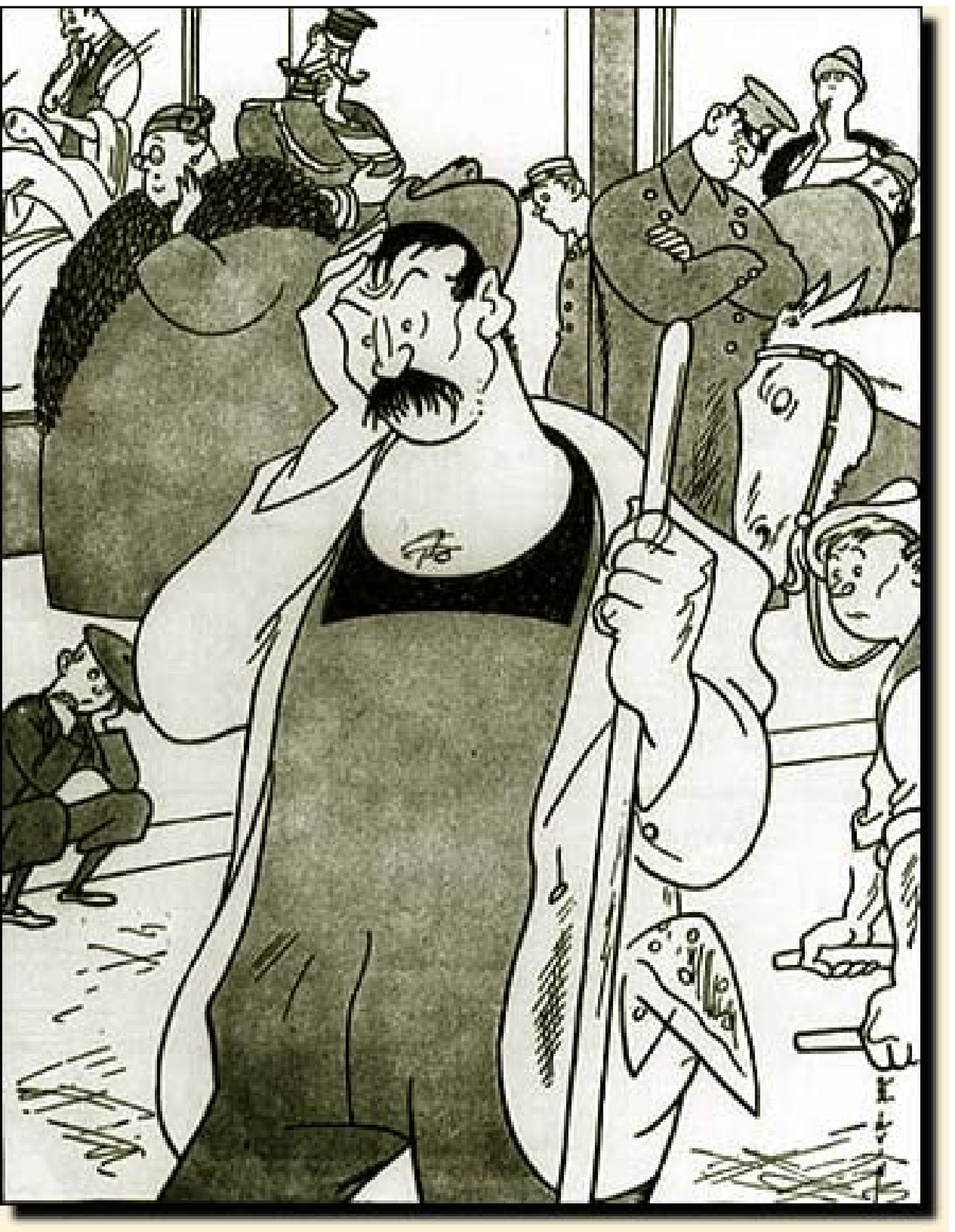}

\begin{quote}
\begin{quote}
``\emph{People slowly accustomed themselves to the idea that the 
physical states of space itself were the final physical 
reality....''}\\ (A. Einstein, 1929)
\end{quote}
\end{quote}
\caption{\label{fig:NewYorker}
Cartoon of 1929 in \emph{The New Yorker} by its first art editor Rea Irvin.}
\end{figure}

That matter-free physical space should have physical properties 
at all seems to be quite against the view of Leibniz, Mach, 
and their modern followers, according to which space is a 
relational concept whose ontological status derives from 
that of the fundamental constituents of matter whose relations 
are considered. 
But at the same time it also seems to be a straightforward 
consequence of modern field theory, according to which 
fundamental fields are directly associated with space (or spacetime) 
rather than any space-filling material substance. Once the latter
view is adopted, there seems to be no good reason to neglect the 
field that describes the geometry of space. This situation 
was frequently and eloquently described by Einstein, who 
empathetically wrote about the difficulties that 
one encounters in attempting to mentally emancipate the notion 
of a field from the idea of a substantial carrier whose physical 
states the field may describe. In doing this, the field describes
the states of space itself, so that space becomes a dynamical 
agent, albeit one to which standard kinematical states of motion
cannot be attributed, as Einstein stressed e.g. in his 1920 
Leiden address ``\"Ather und Relativit\"atstheorie'' 
(\cite{Einstein:CP}\, Vol.\,7, Doc.\,38, pp.\,306-320). 
A famous and amusing cartoon is shown in Figure\,\ref{fig:NewYorker},
whose caption quotes Einstein expressing a view close to that 
of Clifford's.  

\section{Geometrodynamics} 
\label{sec:Geometrodynamics}
\begin{figure}[ht]
\centering
\includegraphics[width=0.80\linewidth]{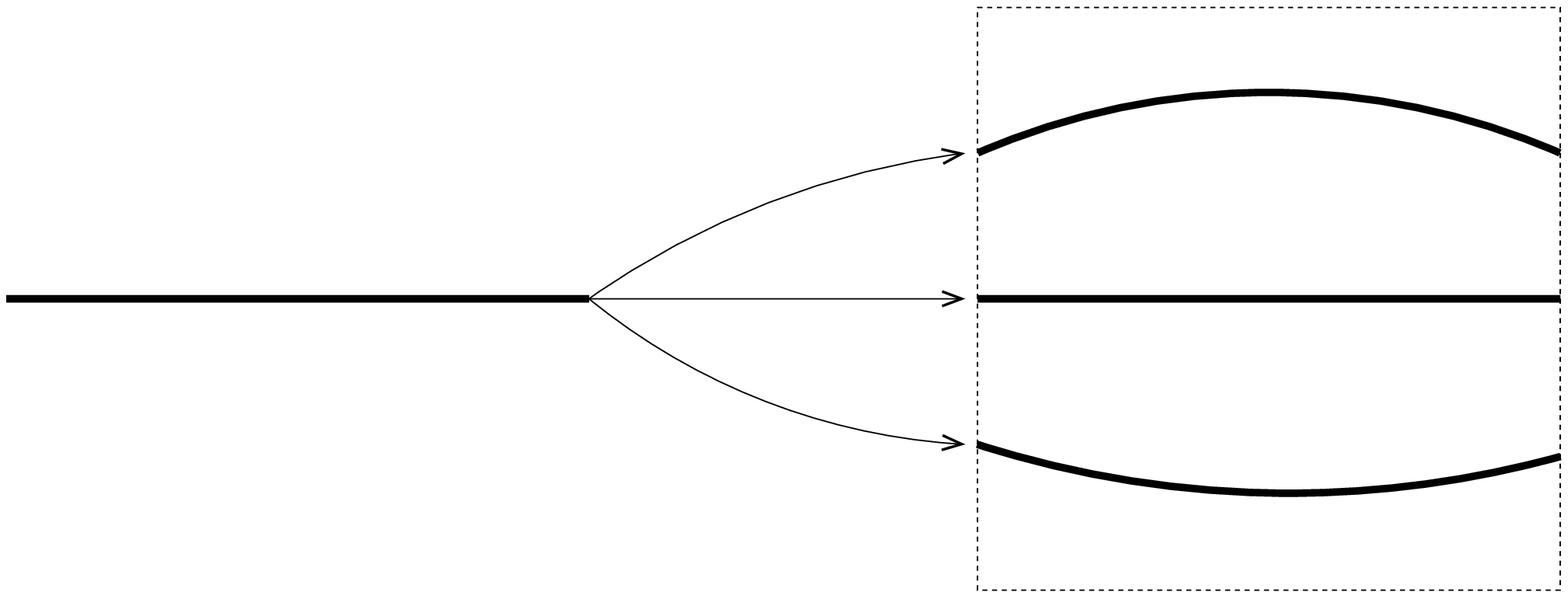}
\put(-98,88){\large $M$}
\put(-225,56){\large $\Sigma$}
\put(-130,76){$\Emb_{1}$}
\put(-130,55){$\Emb_{0}$}
\put(-130,33){$\Emb_{-1}$}
\put(-60,90){$\Sigma_{1}$}
\put(-60,55){$\Sigma_{0}$}
\put(-60,23){$\Sigma_{-1}$}
\caption{\label{fig:Embeddings}
Spacetime $M$ is foliated by a one-parameter family of spacelike 
embeddings of the 3-manifold $\Sigma$. Here the image 
$\Sigma_{1}$ of $\Sigma$ under $\Emb_{t=1}$ lies to the 
future (above) and $\Sigma_{-1}:=\Emb_{t=-1}$ to the past 
(below) of $\Sigma_0:=\Emb_{t=0}(\Sigma)$.}
\end{figure}

The field equations of General Relativity with cosmological
constant $\Lambda$ read ($\kappa=8\pi G/c^4$, where $G$ is 
Newton's constant)
\begin{equation}
\label{eq:EinsteinEquations}
R_{\mu\nu}-\tfrac{1}{2}g_{\mu\nu}R+g_{\mu\nu}\Lambda=\kappa T_{\mu\nu}\,.
\end{equation}
They form a system of ten quasilinear partial differential 
equations for the ten components $g_{\mu\nu}$ of the 
spacetime metric. These equations may be cast into the form 
of evolution equations. More precisely, the system 
(\ref{eq:EinsteinEquations}) may be decomposed into a 
subsystem of four under-determined elliptic equations 
that merely constrain the initial data (the so-called 
`constraints') and a complementary subsystem of six 
under-determined hyperbolic equations that drives the 
evolution. (The under-determination is in both cases a
consequence of diffeomorphism invariance.)
This split is made possible by foliating 
spacetime $M$ into 3-dimensional spacelike leaves 
$\Sigma_t$ via a one-parameter family of embeddings
$\Emb_t:\Sigma\hookrightarrow M$ with images 
$\Emb_t(\Sigma)=\Sigma_t\subset M$; see Fig.\,\ref{fig:Embeddings}. 
The object that undergoes evolution in this picture is the 
3-dimensional Riemannian manifold $(\Sigma,h)$ whose metric 
at time $t$ is $h_t=\Emb^*_tg$, where $g$ is the spacetime 
metric. In this evolutionary picture spacetime appears as 
space's history. 

\subsection{Hypersurface kinematics}
\label{sec:HypersurfaceKinematics}
Let us be more precise on what it means to say that spacetime is 
considered as the trajectory (history) of space. 
Let $\EmbSpace$ denote the space of smooth spacelike embeddings 
$\Sigma\rightarrow M$. We consider a curve 
$\reals\ni t\rightarrow\Emb_t\in\EmbSpace$ corresponding to 
a one-parameter family of smooth embeddings with spacelike 
images. We assume the images $\Emb_t(\Sigma)=:\Sigma_t\subset M$ 
to be mutually disjoint and moreover
$\hat\Emb:\reals\times\Sigma\rightarrow M$, $(t,p)\mapsto\Emb_t(p)$, 
to be an embedding. (It is sometimes found convenient to relax this 
condition, but this is of no importance here). The Lorentz 
manifold $(\reals\times\Sigma,\Emb^*g)$ may now be taken as 
($\Emb$--dependent) representative of $M$ (or at least some open 
part of it) on which the leaves of the above foliation simply 
correspond to the $t=\mathrm{const.}$ hypersurfaces. Let $n$ 
denote a field of normalised timelike vectors normal to these 
leaves. $n$ is unique up to orientation, so that the choice of 
$n$ amounts to picking a `future direction'. 

\begin{figure}[htb]
\centering\includegraphics[width=0.57\linewidth]{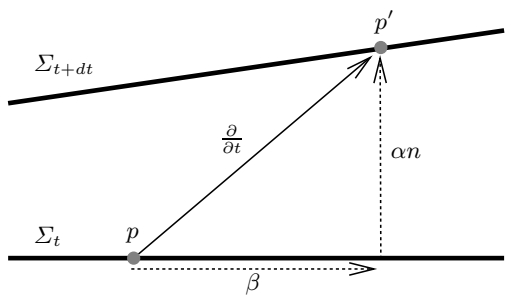}
\put(-180,13){\small $\Sigma_{t}$}
\put(-180,78){\small $\Sigma_{t+dt}$}
\put(-145,15){\small $p$}
\put(-51,94){\small $p'$}
\put(-100,-5){\small $\beta$}
\put(-45,45){\small $\alpha n$}
\put(-110,50){\small $\frac{\partial}{\partial t}$}
\caption{\label{fig:LapseShift}%
For $q\in\Sigma$ the image points $p=\mathcal{E}_t(q)$ and 
$p'=\mathcal{E}_{t+dt}(q)$ are connected by the vector 
$\partial/\partial t\vert_p$ whose components tangential and 
normal to $\Sigma_t$ are $\beta$ (three functions) and $\alpha n$ 
(one function) respectively.}
\end{figure}

The tangent vector $d\Emb_t/dt\vert_{t=0}$ at $\Emb_0\in\EmbSpace$  
corresponds to a vector field over $\Emb_0$ (i.e. section in 
$T(M)\vert_{\Emb_0(\Sigma)}$), given by 
\begin{equation}
\label{eq:LapseShift}
\frac{d\Emb_t(p)}{dt}\Big\vert_{t=0}
=:\frac{\partial}{\partial t}\Big\vert_{\Emb_0(p)}
=\alpha n+\beta
\end{equation}
with components $(\alpha,\beta)$ normal and tangential to $\Sigma_0\subset M$. 
The functions $\alpha$ (one function), usually called the \emph{lapse function},  
and $\beta$ (3 functions), usually called the \emph{shift vector field}, combine 
the four-function worth of arbitrariness in moving the hypersurface $\Sigma$ 
in spacetime; see Fig.\,\ref{fig:LapseShift}.

Conversely, each vector field $V$ on $M$ defines a vector field 
$X(V)$ on $\EmbSpace$, corresponding to the left action of
$\mathrm{Diff}(M)$ on $\EmbSpace$ given by composition. In local 
coordinates $y^\mu$ on $M$ and $x^k$ on $\Sigma$ it can be written as
\begin{equation}
\label{eq:XofV}
X(V)=\int_\Sigma d^3x\,V^\mu(y(x))\frac{\delta}{\delta y^\mu(x)}\,.
\end{equation}
One easily verifies that $X:V\mapsto X(V)$ is a Lie homomorphism:
\begin{equation}
\label{eq:LieHomo}
\bigl[X(V),X(W)\bigr]=X\bigl([V,W]\bigr)\,.
\end{equation}

\begin{figure}[ht]
\centering\includegraphics[width=0.33\linewidth]{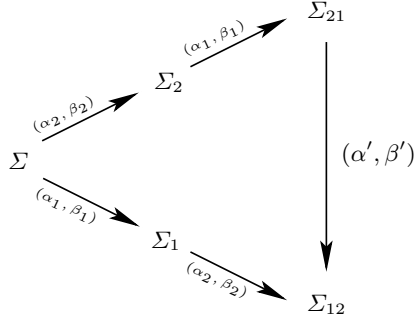}
\put(-123,51){$\Sigma$}
\put(-68,81){$\Sigma_2$}
\put(-69,21){$\Sigma_1$}
\put(-10,108){$\Sigma_{21}$}
\put(-10,-3){$\Sigma_{12}$}
\put(-113,65){\tiny
 \begin{rotate}{25}$(\alpha_2,\beta_2)$\end{rotate}}
\put(-56,94){\tiny
 \begin{rotate}{25}$(\alpha_1,\beta_1)$\end{rotate}}
\put(-113,41){\tiny 
 \begin{rotate}{-25}$(\alpha_1,\beta_1)$\end{rotate}}
\put(-56,13){\tiny 
 \begin{rotate}{-25}$(\alpha_2,\beta_2)$\end{rotate}}
\put(3,54){$(\alpha',\beta')$}
\caption{\label{fig:FigCommDia}%
An (infinitesimal) hypersurface deformation with parameters 
$(\alpha_1,\beta_1)$ that maps $\Sigma\mapsto\Sigma_1$,
followed by one with parameters $(\alpha_2,\beta_2)$ that 
maps $\Sigma_1\mapsto\Sigma_{12}$ differs by one with 
parameters $(\alpha',\beta')$ given by (\ref{eq:X(ab)CommValues})
from that in which the maps with the same parameters are 
composed in the opposite order.}
\end{figure}

Alternatively, decomposing (\ref{eq:XofV}) into normal and tangential 
components with respect to the leaves of the embedding at which the 
tangent-vector field to $\EmbSpace$ is evaluated yields an 
embedding-dependent parametrisation of $X(V)$ in terms of 
$(\alpha,\beta)$,
\begin{equation}
\label{eq:X-alphabeta}
X(\alpha,\beta)=
\int_\Sigma d^3x
\Bigl(\alpha(x)n^\mu[y](x)+\beta^m(x)\partial_m y^\mu(x)
\Bigr)\,\frac{\delta}{\delta y^\mu(x)}\,,
\end{equation}
where $y$ in square brackets indicates the functional dependence 
of $n$ on the embedding. The functional derivatives of $n$ with 
respect to $y$ can be computed (see the Appendix of 
\cite{Teitelboim:1973}) from which the commutator of deformation 
generators follows: 
\begin{equation}
\label{eq:X(ab)Comm}
\bigl[X(\alpha_1,\beta_1)\,,\,X(\alpha_2,\beta_2)\bigr] 
=\,-\,X(\alpha',\beta')\,,
\end{equation}
where 
\begin{subequations}
\label{eq:X(ab)CommValues}
\begin{eqnarray}
\label{eq:X(ab)Comm-b}
&\alpha' &\,=
\beta_1(\alpha_2)-\beta_2(\alpha_1)\,,\\
\label{eq:X(ab)Comm-c}
&\beta' &\,=
[\beta_1,\beta_2]+\sigma\alpha_1\,\mathrm{grad}_h(\alpha_2)
     -\sigma\alpha_2\mathrm{grad}_h\,(\alpha_1)\,.
\end{eqnarray}
\end{subequations}
Here we left open whether spacetime $M$ is Lorentzian ($\sigma=1$) 
or Euclidean ($\sigma=-1$), just in order to keep track how the 
signature of spacetime, $(-\sigma,+,+,+)$, enters. Note that 
the $h$-dependent gradient field for the scalar function $\alpha$ is 
given by $\mathrm{grad}_h(\alpha)=(h^{ab}\partial_b\alpha)\partial_a$. 
The geometric idea behind (\ref{eq:X(ab)CommValues}) is summarised 
in Figure\,\ref{fig:FigCommDia}. 

\subsection{Hamiltonian geometrodynamics}
The idea of Hamiltonian Geometrodynamics is to realise these 
relations in terms of a Hamiltonian system on the phase space of 
physical fields. The most simple case is that where the latter 
merely include the spatial metric $h$ on $\Sigma$, so that the 
phase space is the cotangent bundle $T^*\Riem$ over $\Riem$.
One then seeks a correspondence that associates to each pair 
$(\alpha,\beta)$ of lapse and shift a real-valued function on 
phase space: 
\begin{equation}
\label{eq:PhaseSpaceDist}
(\alpha,\beta)\mapsto\bigl(
H(\alpha,\beta)\,:\,T^*\Riem\rightarrow\reals\bigr)\,,
\end{equation}
where
\begin{equation}
\label{eq:NormalDefCorr}
H(\alpha,\beta)[h,\pi]:=
\int_\Sigma d^3x\bigl(
\alpha(x)\Hcal[h,\pi](x)+h_{ab}(x)\beta^a(x)\Dcal^b[h,\pi](x)\bigr)\,,
\end{equation}
with integrands $\Hcal[h,\pi](x)$ and $\Dcal^b[h,\pi](x)$ yet to 
be determined. $H$ should be regarded as distribution (here the test 
functions are $\alpha$ and $\beta^a$) with values in real-valued 
functions on $T^*\Riem$. Now, the essential requirement is that 
the Poisson brackets between the $H(\alpha,\beta)$ are, up to a 
minus sign,\footnote{Due to the standard convention that the 
Hamiltonian action being defined as a \emph{left} action, whereas 
the Lie bracket on a group is defined by the commutator of 
left-invariant vector fields which generate \emph{right} 
translations.} as in (\ref{eq:X(ab)Comm}-\ref{eq:X(ab)CommValues}): 
\begin{equation}
\label{eq:H(ab)Comm}
\bigl\{H(\alpha_1,\beta_1)\,,\,H(\alpha_2,\beta_2)\bigr\}
=H(\alpha',\beta')\,.
\end{equation}

For the integration of canonical initial data $(h,\pi)$ with 
Hamiltonian $H(\alpha,\beta)$ we need to specify by hand a 
one-parameter (representing parameter time $t$) family of 
lapse functions $t\mapsto \alpha_t(x)=\alpha(t,x)$ and shift 
vector fields $t\mapsto \beta_t(x)=\beta(t,x)$. It is now 
clear that this freedom just corresponds to the freedom to 
foliate the spacetime to be constructed. The Hamiltonian 
equations of motion contain only the unknown functions 
$h_t(x)=h(t,x)$ and $\pi_t(x)=\pi(t,x)$ and should be regarded 
as evolution equations (in terms of parameter time $t$) 
for the one-parameter families of tensor fields $t\mapsto h_t$
and $t\mapsto \pi_t$. Once the integration is performed, the 
solution gives rise to solution of Einstein's equation:
If $\beta_t^\flat$ is the one-form field corresponding to the 
vector field $\beta_t$ via $h_t$, 
i.e. $\beta_t^\flat:=h_t(\beta_t,\cdot)$, then the 
Lorentzian metric that satisfies Einstein's equation on the 
manifold $I\times\Sigma$, where $I$ is the interval on 
the real line in which the parameter $t$ takes its values, 
is given by 
\begin{equation}
\label{eq:SolMetricEinst}
g=-\bigl(\alpha_t^2-h_t(\beta_t,\beta_t)\bigr)dt\otimes dt
+\beta_t^\flat\otimes dt+dt\otimes\beta_t^\flat+h_t\,.
\end{equation}
However, this integration may not start from any arbitrary
set of initial data $(h,\pi)$. The data themselves need to satisfy 
a system of (under-determined elliptic) partial differential
equations, the so-called `constraints'. The reason for their 
existence as well as their analytic form will be explained in 
the next subsections.

\subsection{Why constraints}
From (\ref{eq:H(ab)Comm}) alone follows a remarkable uniqueness 
result as regards the analytical structure of $H(\alpha,\beta)$ 
as functional of $(h,\pi)$. Before stating it with all its hypotheses, 
we show why the constraints $\Hcal[h,\pi]=0$ and $\Dcal^b[h,\pi]=0$ 
must be imposed. 

Consider the set of smooth real-valued functions on
phase space, $F:T^*\Riem\rightarrow\reals$. They are 
acted upon by all $H(\alpha,\beta)$ via Poisson bracketing:
$F\mapsto\bigl\{F,H(\alpha,\beta)\bigr\}$. This defines a
map from $(\alpha,\beta)$ into the derivations of phase-space
functions. We require this map to also respect the 
commutation relation (\ref{eq:H(ab)Comm}), that is, we require 
\begin{equation}
\label{eq:Der(ab)Comm-1}
\bigl\{\bigl\{F,H(\alpha_1,\beta_1)\bigr\},H(\alpha_2,\beta_2)\bigr\}-
\bigl\{\bigl\{F,H(\alpha_2,\beta_2)\bigr\},H(\alpha_1,\beta_1)\bigr\}
=\bigl\{F,H\bigr\}(\alpha',\beta')\,.
\end{equation}
The crucial and somewhat subtle point to be observed here is the 
following: Up to now  
the parameters $(\alpha_1,\beta_1)$ and $(\alpha_2,\beta_2)$ were 
considered as given functions of $x\in\Sigma$, independent of the 
fields $h$ and $\pi$, i.e. independent of the point of phase 
space. However, from (\ref{eq:X(ab)Comm-c}) we see that $\beta'(x)$ 
does depend on $h(x)$. This dependence should not give rise to extra terms 
$\propto\{F,\alpha'\}$ in the Poisson bracket, for, otherwise, the
extra terms would prevent the map 
$(\alpha,\beta)\mapsto\bigl\{-,H(\alpha,\beta)\bigr\}$
from being a homomorphism from the algebraic structure of 
hypersurface deformations into the derivations of phase-space 
functions. This is necessary in order to interpret 
$\bigl\{-,H(\alpha,\beta)\bigr\}$ as a generator (on phase-space 
functions) of a \emph{spacetime} evolution corresponding to a normal 
lapse $\alpha$ and tangential shift $\beta$. In other words, the 
evolution of observables from an initial hypersurface $\Sigma_i$ 
to a final hypersurface $\Sigma_f$ must be independent of the 
intermediate foliation (`integrability' or `path independence'
~\cite{Teitelboim:1973,Hojman.etal:1973,Hojman.etal:1976}).
Therefore we placed the parameters $(\alpha',\beta')$ outside the 
Poisson bracket on the right-hand side of (\ref{eq:Der(ab)Comm-1}), 
to indicate that no differentiation with respect to $h,\pi$ should 
act on them. 

To see that this requirement implies the constraints, rewrite 
the left-hand side of (\ref{eq:Der(ab)Comm-1}) in the form
\begin{equation}
\label{eq:Der(ab)Comm-2}
\begin{split}
& \bigl\{\bigl\{F,H(\alpha_1,\beta_1)\bigr\},H(\alpha_2,\beta_2)\bigr\}-
\bigl\{\bigl\{F,H(\alpha_2,\beta_2)\bigr\},H(\alpha_1,\beta_1)\bigr\}\\
&\quad=\,\bigl\{F,\bigl\{H(\alpha_1,\beta_1),H(\alpha_2,\beta_2)\bigr\}\bigr\}\\
&\quad=\,\bigl\{F,H(\alpha',\beta')\bigr\}\\
&\quad=\,\bigl\{F,H\bigr\}(\alpha',\beta')
  +H\bigl(\{F,\alpha'\}\,,\,\{F,\beta'\}\bigr)\,,
\end{split}
\end{equation}
where the first equality follows from the Jacobi identity, the second 
from (\ref{eq:H(ab)Comm}), and the third from the Leibniz rule. 
Hence the requirement (\ref{eq:Der(ab)Comm-1}) is equivalent to 
\begin{equation}
\label{eq:Der(ab)Comm-3}
H\bigl(\{F,\alpha'\}\,,\,\{F,\beta'\}\bigr)=0
\end{equation}
for all phase-space functions $F$ to be considered and all 
$\alpha',\beta'$ of the form (\ref{eq:X(ab)CommValues}). 
Since only $\beta'$ depends on phase space, more precisely on $h$, 
this implies the vanishing of the phase-space functions 
$H\bigl(0,\{F,\beta'\}\bigr)$ for all $F$ and all $\beta'$ 
of the form (\ref{eq:X(ab)Comm-c}). This can be shown to 
imply $H(0,\beta)=0$, i.e. $\Dcal[h,\pi]=0$. Now, 
in turn, for this to be preserved under all evolutions we need 
$\bigl\{H(\alpha,\tilde \beta),H(0,\beta)\bigr\}=0$, and hence 
in particular $\bigl\{H(\alpha,0),H(0,\beta)\bigr\}=0$ for all 
$\alpha,\beta$, which implies $H(\alpha,0)=0$, i.e. $\Hcal[h,\pi]=0$. 
So we see that the constraints indeed follow from the required 
integrability condition. 

Sometimes the constraints $H(\alpha,\beta)=0$ are split into 
the \emph{Hamiltonian (or scalar) constraints}, $H(\alpha,0)=0$, and 
the \emph{diffeomorphisms (or vector) constraints}, $H(0,\beta)=0$. 
The relations (\ref{eq:H(ab)Comm}) with (\ref{eq:X(ab)CommValues})
then show that the vector constraints form a 
Lie-subalgebra which, because of $\{H(0,\beta),H(\alpha,0)\}
=H\bigl(\beta(\alpha),0\bigr)\ne H(0,\beta')$, is not an 
ideal. This means that the Hamiltonian vector fields for the 
scalar constraints are not tangent to the surface of 
vanishing  vector constraints, except where it intersects the 
surface of vanishing scalar constraints. This implies that the 
scalar constraints do not act on the solution space for the 
vector constraints, so that one simply cannot first reduce the 
vector constraints and then, on the solutions of that, search  
for solutions to the scalar constraints. 

\subsection{Uniqueness of Einstein's geometrodynamics}
It is sometimes stated that the relations (\ref{eq:H(ab)Comm})
together with (\ref{eq:X(ab)CommValues}) determine the 
function $H(\alpha,\beta):T^*\Riem\rightarrow\reals$, i.e. 
the integrands $\Hcal[h,\pi]$ and $\Dcal[h,\pi]$, uniquely 
up to two free parameters, which may be identified with the 
gravitational and the cosmological constants. This is a 
mathematical overstatement if read literally, since the 
result can only be shown if certain additional assumptions 
are made concerning the action of $H(\alpha,\beta)$ on the 
basic variables $h$ and $\pi$. The uniqueness result then 
obtained is still remarkable. 

The first such assumption concerns the intended (`semantic' or
`physical') meaning of $H(0,\beta)$, namely that the action
of $H(0,\beta)\}$ on $h$ or $\pi$ is that of an infinitesimal 
spatial diffeomorphism of $\Sigma$. Hence it should be the 
spatial Lie derivative, $L_\beta$, applied to $h$ or $\pi$. 
It then follows from the general Hamiltonian theory that 
$H(0,\beta)$ is given by the \emph{momentum map} that maps the 
vector field $\beta$ (viewed as element of the Lie algebra 
of the group of spatial diffeomorphisms) into the function 
on phase space given by the contraction of the momentum 
with the $\beta$-induced vector field $h\rightarrow L_\beta h$ 
on $\Riem$:
\begin{equation}
\label{eq:MomentiumMap}
H(0,\beta)=\int_\Sigma d^3x\,
\pi^{ab}(L_\beta h)_{ab}
=-2\int_\Sigma d^3x (\nabla_a\pi^{ab})h_{bc}\beta^c\,.
\end{equation}
Comparison with (\ref{eq:NormalDefCorr}) yields
\begin{equation}
\label{eq:DiffConstrFunct}
\Dcal^b[h,\pi]=-2\nabla_a\pi^{ab}\,.
\end{equation}

The second assumption concerns the intended (`semantic' or
`physical') meaning of $H(\alpha,0)$, namely that 
$\{-,H(\alpha,0)\}$ acting on $h$ or $\pi$ is that of an 
infinitesimal `timelike' diffeomorphism of $M$ normal to 
the leaves $\Emb_t(\Sigma)$. If $M$ were given, it is 
easy to prove that we would have $L_{\alpha n}h=2\alpha\,K$, 
where $n$ is the timelike field of normals to the leaves 
$\Emb_t(\Sigma)$ and $K$ is their extrinsic curvature.
Hence one requires 
\begin{equation}
\label{eq:Anticip}
\{h,H(\alpha,0)\}=2\alpha\,K\,.
\end{equation}
Note that both sides are symmetric covariant tensor fields over
$\Sigma$. The important fact to be observed here is that $\alpha$
appears without differentiation. This means that $H(\alpha,0)$ 
is an ultralocal functional of $\pi$, which is further assumed to be 
a polynomial. Note that at this moment we do not assume any 
definite relation between $\pi$ and $K$. Rather, this relation is a 
consequence of (\ref{eq:Anticip}) once the analytic form of 
$H(\alpha,0)$ is determined. 

The Hamiltonian evolution so obtained is precisely that of 
General Relativity (without matter) with two free parameters,
which may be identified with the gravitational 
constant $\kappa=8\pi G/c^4$ and the cosmological constant 
$\Lambda$. The proof of the theorem is given in \cite{Kuchar:1973},
which improves on earlier versions \cite{Teitelboim:1973,Hojman.etal:1976} 
in that the latter assume in addition that $\Hcal[h,\pi]$ be 
an even function of $\pi$, corresponding to the requirement 
of time reversibility of the generated evolution.  This was 
overcome in \cite{Kuchar:1973} by the clever move to write 
the condition set by $\{H(\alpha_1,0),H(\alpha_2,0)\}=H(0,\beta')$ 
(the right-hand side being already known) on $H(\alpha,0)$ in 
terms of the corresponding Lagrangian 
functional $L$, which is then immediately seen to turn into a 
condition which is \emph{linear} in $L$, so that terms with 
even powers in velocity decouple form those with odd powers. 
There is a slight topological subtlety remaining which is further 
discussed in \cite{Giulini:2009a}. The two points which are 
important for us here are:
\begin{enumerate}
\item
\emph{
The dynamics of the gravitational field as given by 
Einstein's equations can be fully understood in term 
of the constraints.}  
\item
\emph{
Modulo some technical assumptions spelled out above, the constraints for pure gravity 
follow from the kinematical relation (\ref{eq:H(ab)Comm}) with (\ref{eq:X(ab)CommValues}),
once one specifies and gravitational phase space to be 
$T^*\Riem$, i.e. the gravitational configuration space to be 
$\Riem$.} 
\end{enumerate}

\subsection{What the constraints look like}
\label{sec:ConstrainedDynamics}
Rather than writing down the constraints in terms of $h$ and $\pi$, 
we shall use the simple relation between $\pi$ and $K$ that follows 
from (\ref{eq:Anticip}) for given $H(\alpha,0)$, the reason being that
$K$ has the simple interpretation as extrinsic curvature (also called
second fundamental form) of the images of $\Sigma$ in $M$, which is 
rather intuitive. From the determination of $H(\alpha,0)$ the 
$h$-dependent relation between $\pi$ and $K$, in terms of components,
turns out to be  
\begin{equation}
\label{eq:RelPiExtCurv}
\pi^{ab}=\sqrt{\det\{h_{nm}\}}\,
h^{ac}h^{bd}\bigl(K_{cd}-h_{cd}h^{ij}K_{ij})\,.
\end{equation}
In terms of $h$ and $K$ the constraints then assume the form 
\begin{subequations}
\label{eq:Constraints}
\begin{eqnarray}
\label{eq:Constraints-a}
\bigl(h^{ac}h^{bd}\,K_{ab}K_{cd}-(h^{ab}K_{ab})^2\bigr)
-\bigl(R(h)-2\Lambda\bigr)&=-&(2\kappa)\rho\,,\\
\label{eq:Constraints-b}
\nabla_b\bigl(K^{ab}-h^{ab}\,h^{nm}K_{nm}\bigr)
&=&(c\kappa)\,j^a\,,
\end{eqnarray}
\end{subequations}
The right-hand sides of both equations (\ref{eq:Constraints}) 
are zero in the matter free (vacuum) case which we consider here. 
But we think it is instructive to know what it will be in the 
presence of matter: Here $\rho$ and $j$ represent the matter's 
energy and momentum densities on $\Sigma$ respectively.\footnote{%
Recall that the symmetry of the energy-momentum tensor for 
the matter implies that the momentum density is $c^{-2}$ times 
the energy current density (energy per unit surface area and 
unit time).}%
Moreover, $R(h)$ is the Ricci scalar for $h$, $\Lambda$ is 
the cosmological constant, and $\kappa=8\pi G/c^4$ as in 
(\ref{eq:EinsteinEquations}). 

The first bracket on the left-hand side of (\ref{eq:Constraints-a})
contains an $h$-dependent  bilinear form in $K$. It can be seen 
as the kinetic term in the Hamiltonian of the gravitational field. 
Usually the kinetic term is positive definite, but this time it 
is not! Hence we wish to understand this bilinear form in more 
detail. In particular: Under what conditions on $K$ is 
it positive or negative definite? This can be answered in terms 
of the eigenvalues of $K$. To make this precise, let $\tilde K$ 
be the endomorphism field which is obtained from $K$ by raising one 
index (which one does not matter due to symmetry) using $h$. 
We may now unambiguously speak of the eigenvalues of $\tilde K$,
a triple for each space point. Each triple we collect in an 
eigenvalue vector $\vec\lambda\in\reals^3$. In terms of $\tilde K$ 
the bilinear form reads $\tr(\tilde K^2)-(\tr(\tilde K))^2$,
which equals $\Vert\vec \lambda\Vert^2-(\vec\lambda\cdot\vec d)^2$
in terms of $\vec\lambda$. Here the dot product and the norm are 
the usual ones in $\reals^3$ and $\vec d$ is the `diagonal vector'
with unit entries $(1,1,1)$. Hence the bilinear form is positive 
definite iff\footnote{In this article we use `iff' as abbreviation 
for `if and only if'.} the modulus of the cosine of the angle between $
\vec\lambda$ and $\vec d$ is less than $1/\sqrt{3}$ and negative 
definite iff it is greater than $1/\sqrt{3}$. In other words,
the bilinear form is negative definite on those $\tilde K$ whose 
eigenvalue vector lies in the interior of a double cone whose 
vertex is the origin, whose symmetry axis is the `diagonal' generated 
by $\vec d$, and whose opening angle (angle between symmetry axis 
and boundary) is $\arccos(1/\sqrt{3})\approx 54.7^{\circ}$.
Note that this opening angle is just the one at which the 
boundary of each cone just contains all three positive or 
negative coordinate half-axes. It properly contains the 
maximal cones contained in the positive and negative octants,
whose opening angle is $\arccos(\sqrt{2/3})\approx 35.3^{\circ}$.%
\footnote{The maximal cone touches the three 2-planes $\lambda_i=0$ at the 
bisecting lines $\lambda_j=\lambda_k$, where $i,j,k$ is any of 
the three cyclic permutations of $1,2,3$. Hence the cosine of
the opening angle is the scalar product between 
$(1,1,1)\sqrt{3}$ and, say, $(1,1,0)\sqrt{2}$, which is 
$\sqrt{2/3}$.}
Hence strictly positive or strictly negative eigenvalues of 
$\tilde K$ imply a negative definite value of the bilinear 
form, but the converse is not true.%
\footnote{Had we done the very same analysis in terms of $\pi$ 
rather than $K$ we would have found that in eigenvalue space (now 
of the endomorphism $\tilde\pi$) the opening angle of the cone 
inside which the bilinear form is negative definite and outside which 
it is positive definite is now \emph{precisely} the maximal one 
$\arccos(\sqrt{2/3})$ (see previous footnote). Indeed, rewriting 
the bilinear form in terms of $\pi$ using (\ref{eq:RelPiExtCurv}), 
it is positively proportional to 
$\tr(\tilde \pi^2)-\frac{1}{2}(\tr(\tilde \pi))^2$. It is the 
one-half in front of the second term that causes this interesting 
coincidence.}

The preceding discussion shows that the bilinear form is not of a 
definite nature. In fact, it is a $(1+5)$ -- dimensional Lorentzian 
metric on the six-dimensional space of positive definite bilinear 
forms over a real 3-dimensional vector space (the tangent space to 
$\Sigma$), which is known as the DeWitt metric since DeWitt's
seminal paper \cite{DeWittQTGI:1967} on canonical quantum 
gravity.  Parametrising it by $h_{ab}$ or $(\tau,r_{ab})$ it can 
be written as     
\begin{subequations}
\label{eq:DWmetric}
\begin{equation}
\label{eq:DWmetric-a}
G^{ab\,cd}dh_{ab}\otimes dh_{cd}
=\,-\, (32/3)\,d\tau\otimes d\tau\,+\,\tau^2\
\tr(r^{-1}dr\otimes r^{-1}dr)\,,
\end{equation}
where 
\begin{equation}
\label{eq:DWmetric-b}
r_{ab}:=[\det(h)]^{-1/3}h_{ab},\ \
\tau:=[\det(h)]^{1/4}\,,
\end{equation}
and 
\begin{equation}
\label{eq:DWmetric-c}
G^{ab\,cd}=\tfrac{1}{2}\sqrt{\det(h)}
\bigl(h^{ac}h^{bd}+h^{ad}h^{bc}-2 h^{ab}h^{cd}\bigr)\,.
\end{equation}
\end{subequations}
The form (\ref{eq:DWmetric-a}) clearly reveals 
it geometric meaning as a warped-product metric of
`cosmological type' on the manifold $\reals\times SL(3,\reals)/SO(3)$, 
where the five-dimensional homogeneous space $SL(3,\reals)/SO(3)$, 
parametrised by $r_{ab}$, carries its left invariant metric 
$\tr(r^{-1}dr\otimes r^{-1}dr)=r^{ac}r^{bd}\,dr_{ab}\otimes dr_{cd}$. 

This pointwise Lorentzian metric induces a metric on the 
infinite-dimensional manifold $\Riem$, known as Wheeler-DeWitt
metric, through 
\begin{equation}
\label{eq:WDWmetric}
\G(k,\ell)=\int_\Sigma d^3x\, G^{qb\,cd} k_{ab}\ell_{cd}\,
\end{equation}
where the tensor fields $k$ and $\ell$ are now considered as 
tangent vectors at $h\in\Riem$. See \cite{Giulini:2009a} and 
references therein for a recent review on geometric aspects 
associated with this metric and its r\^ole in Geometrodynamics.

To end this brief sketch of Geometrodynamics let me just stress 
its (admittedly somewhat crude) analogy to relativistic point
mechanics. The latter takes place in Minkowski space which is 
endowed with an absolute (i.e. non dynamical) geometry through 
the Minkowski metric. Here the configuration space is $\Riem$,
which is also endowed with an absolute geometry through the 
Wheeler-DeWitt metric, although it is not true that the 
Einstein equations correspond to geodesic motion in it. 
However, the deviation from geodesic motion derives from a 
force that corresponds to a vector field on $\Riem$ given 
by $-2(R_{ab}-\tfrac{1}{4}h_{ab}R)$, where $R_{ab}$ and $R$
are the Ricci tensor and scalar for $h$~\cite{Giulini:1995c}
respectively. 

\subsection{Vacuum Data}
\label{sec:VacuumData}
Following Clifford's dictum, we shall in the following be interested 
in vacuum data, that is data $(h,K)$ that satisfy (\ref{eq:Constraints}) 
for vanishing right-hand sides. Upon evolution these give rise to solutions
$g_{\mu\nu}$ to Einstein's equations (\ref{eq:EinsteinEquations})
for $T_{\mu\nu}=0$. 

An important non-trivial observation is that the system 
(\ref{eq:Constraints}) does not impose any topological obstruction 
on $\Sigma$. That means that for any topological 3-manifold 
$\Sigma$ there are data $(h,K)$ that satisfy 
(\ref{eq:Constraints}) with vanishing right-hand side. 
This result can be understood as an immediate consequence of a 
famous theorem proved in~\cite{Kazdan.Warner:1975}, that states 
that \emph{any} smooth function $f:\Sigma\rightarrow\reals$ which 
is negative somewhere can be the scalar curvature for some 
Riemannian metric. Given that strong result, we may indeed always 
solve (\ref{eq:Constraints}) for $\rho=0$ and $j=0$ as follows: 
First we make the Ansatz $K=\alpha h$ for some constant $\alpha$ 
and some $h\in\Riem$. This solves (\ref{eq:Constraints-b}), whatever 
$\alpha,h$ will be. Given the space-time interpretation of $K$
as extrinsic curvature, this means that the initial $\Sigma$ 
will be a totally umbillic hypersurface in the spacetime $M$ that 
is going to evolve from the data. Next we solve (\ref{eq:Constraints-a}) 
by fixing $\alpha$ so that $\alpha^2>\Lambda/3$ and then choosing 
$h$ so that $R(h)=2\Lambda-6\alpha^2$, which is possible by the 
result just cited because the right-hand side is negative by 
construction. 
 
Simple but nevertheless very useful examples of vacuum data 
are provided by time-symmetric conformally flat ones. 
Time symmetry means that the initial `velocity' of $h$ vanishes,
and hence that $K=0$, so that~(\ref{eq:Constraints-b}) is 
already satisfied. A vanishing extrinsic curvature is equivalent 
to saying that the hypersurface is totally geodesic, meaning 
that a geodesic in spacetime that initially starts on and 
tangent to $\Sigma\subset M$ will remain within $\Sigma$.
This is to be expected since motions with vanishing initial 
velocity should be time-reflection symmetric, which here 
would imply the existence of an isometry of $M$ (the history 
of space) that exchanges both sides of $\Sigma$ in $M$
and leaves $\Sigma$ pointwise fixed. However, a fixed-point
set of an isometry is always totally geodesic, for, if it were 
not, a geodesic starting on and tangent to $\Sigma$ but taking 
off $\Sigma$ eventually would be mapped by the isometry to 
a different geodesic with the same initial conditions, which 
contradicts the uniqueness theorem for solutions of the 
geodesic equation.   

As time symmetry implies $K=0$, we have automatically solved 
(\ref{eq:Constraints-b}) for $j=0$. That $h$ be conformally 
flat means that we may write $h=\phi^4\,\delta$, where $\delta$ 
is the flat euclidean metric on $\Sigma$ and 
$\phi:\Sigma\rightarrow\reals_+$ is a positive real-valued 
function. The remaining constraint (\ref{eq:Constraints-a})
for $\rho=0$ then simply reduces to Laplace's equation for 
$\phi$:
\begin{equation}
\label{eq:LaplaceEq}
\Delta\phi=0\,,
\end{equation}
where $\Delta$ denotes the Laplace operator with respect to 
the flat metric $\delta$. 

Usually one seeks solutions so that $(\Sigma,h)$ is a manifold 
with a finite number of asymptotically flat ends. One such end is 
then associated with `spatial infinity', which really just means 
that the solution under consideration represents a quasi isolated
lump of geometry with a sufficiently large (compared to its own 
dimension) almost flat transition region to the ambient universe.
According to Arnowitt, Deser, and Misner (see their review 
\cite{Arnowitt.Deser.Misner:1962}) we can associate a 
(active gravitational) mass to each such end, which is defined 
as a limit of a flux integrals over 2-spheres pushed into the 
asymptotic region of the end in question. If the mass is measured 
in geometric units (i.e. it has the physical dimension of a 
length and is converted to a mass in ordinary units by 
multiplication with $c^2/G$), it is given by 
\begin{equation}
\label{eq:DefADMMass}
m:=\lim_{R\rightarrow\infty}
\left\{
\frac{1}{16\pi}\int_{S_R}(\partial_jh_{ij}-\partial_ih_{jj})n^i\,d\Omega
\right\}\,,
\end{equation}
where $S_R\subset\Sigma$ is a 2-sphere of radius $R$, 
outward-pointing normal $n$, and surface measure $d\Omega$. 
For later use we note in passing that if the asymptotically 
flat spacetime is globally stationary (i.e. admits a timelike 
Killing field $K$), the overall mass can also be written 
in the following simple form, known as `Komar 
integral'~\cite{Komar:1959}:
\begin{equation}
\label{eq:DefKomarMass}
m:=\lim_{R\rightarrow\infty}
\left\{
\frac{-1}{8\pi}\int_{S_R}\star dK^\flat
\right\}\,,
\end{equation}
where $d$ denotes the exterior differential on spacetime, 
$K^\flat:=g(K,\,\cdot\,)$ the one-form corresponding to $K$ 
under the spacetime metric $g$ (lowering the index) and 
$\star$ is the Hodge-duality map. A similar expression exists 
for the overall angular momentum of a rotationally symmetric 
spacetime, as we shall see later.   

The celebrated `positive-mass theorem' states in the vacuum 
case that for any Riemannian metric $h$ which satisfies the 
constraints (\ref{eq:Constraints}) with $\rho=0$ and $j=0$ 
for some $K$ has $m\geq 0$ for each asymptotically flat end.%
\footnote{Note that the definition of the ADM mass
(\ref{eq:DefADMMass}) just depends on the Riemannian metric $h$
and is independent of $K$. But for the theorem to hold it is 
essential to require that $h$ is such that there exists a $K$
so that $(h,K)$ satisfy the constraints. It is easy to write 
down metrics $h$ with negative mass: Take e.g. 
(\ref{eq:SinglePoleMetric}) with negative $m$ for $r>r_*>m/2$, 
smoothly interpolated within $m/2<r<r_*$ to, say, the flat metric in 
$r<m/2$. The positive mass theorem implies that for such a metric 
no $K$ can be found so that $(h,K)$ satisfy the constraints.}
Moreover, $m=0$ iff $(\Sigma,h)$ is a spatial slice 
through Minkowski space. This already implies that the mass
must be strictly positive if $\Sigma$ is topologically different 
from $\reals^3$: Non-trivial topology implies non-zero positive
mass! This is supported by the generalisation of the Penrose-Hawking
singularity theorems due to Gannon~\cite{Gannon:1975}, which basically 
states that the geometric hypothesis of the existence of closed trapped 
surfaces in $\Sigma$ in the former may be replaced by the purely 
topological hypothesis of $\Sigma$ not being simply connected.
This is our first example in GR of how attributes of matter (here 
mass) arise from pure geometry/topology.

\subsection{Solution strategies}
\label{sec:Solution strategies}
A variety of methods exist to construct interesting solutions to 
(\ref{eq:LaplaceEq}). On of them is the `method of images' known 
from electrostatics~\cite{Misner:1963}. It is based on the conformal 
properties of the Laplace operator, which are as follows: 
Let $\Sigma=\reals^3-\{\vec x_0\}$
and $\delta$ its usual flat metric. Consider a sphere $S_0$ of 
radius $r_0$ centred at $\vec x_0$. The `inversion at $S_0$',
denoted by $I_{(\vec x_0,r_0)}$, is a diffeomorphism of $\Sigma$ that 
interchanges the exterior and the interior of $S_0\subset\Sigma$ 
and leaves $S_a$ pointwise fixed. In spherical polar coordinates    
centred at $\vec x_0$ it takes the simple form 
\begin{subequations}
\label{eq:Inversion}
\begin{equation}
\label{eq:Inversion-a}
I_{(\vec x_0,r_0)}(r,\theta,\varphi)=(r_0^2/r,\theta,\varphi)\,.
\end{equation}
There is a variant of this map that results from an additional 
antipodal reflection in the 2-spheres so that no fixed points 
exist:
\begin{equation}
\label{eq:Inversion-b}
I'_{(\vec x_0,r_0)}(r,\theta,\varphi)=(r_0^2/r,\pi-\theta,\varphi+\pi)\,.
\end{equation}
\end{subequations}
Associated to each of these self-maps of $\Sigma$ are self-maps 
$J_{(\vec x_0,r_0)}$ and $J'_{(\vec x_0,r_0)}$ of the set of smooth 
real-valued functions on $\Sigma$, given by 
\begin{equation}
\label{eq:InversionFunction} 
J_{(\vec x_0,r_0)}(f)=(r_0/r)\,(f\circ I_{(\vec x_0,r_0)})
\end{equation}
and likewise with $I'_{(\vec x_0,r_0)}$ exchanged for $I_{(\vec x_0,r_0)}$ 
on the right-hand side in case of $J'_{(\vec x_0,r_0)}$. Now, the point 
is that these maps obey the following simple composition laws with the 
Laplace operator (considered as self-map of the set of smooth functions 
on $\Sigma$): 
\begin{equation}
\label{eq:LaplafveInversion}
\Delta\circ J_{(\vec x_0,r_0)}= (r_0/r)^4\,J_{(\vec x_0,r_0)}\circ\Delta
\end{equation}    
and likewise with $J'_{(\vec x_0,r_0)}$ replacing $J_{(\vec x_0,r_0)}$.
In particular, the last equation implies that $J_{(\vec x_0,r_0)}$
and $J'_{(\vec x_0,r_0)}$ map harmonic functions (i.e. functions 
$\phi$ satisfying $\Delta\phi=0$) on $\Sigma$ to harmonic functions
on $\Sigma$. Note that $\Sigma$ did not include the point $\vec x_0$
at which the sphere of inversion was centred. It is clear from 
(\ref{eq:InversionFunction}) that the maps $J_{(\vec x_0,r_0)}$
will change the singular behaviour of the functions at $\vec x_0$. 
For example, the image of the constant function $f\equiv 1$ under 
either $J_{(\vec x_0,r_0)}$ or $J_{(\vec x_0,r_0)}$ is just 
the function $\vec x\mapsto r_0/\Vert\vec x-\vec x_0\Vert$, i.e. 
a pole of strength $r_0$ at $\vec x_0$. Iterating once more, a pole 
of strength $a$ located at $\vec a$ is mapped via $J_{(\vec x_0,r_0)}$
resp. $J'_{(\vec x_0,r_0)}$ to a pole of strength 
$a/\Vert\vec x_0-\vec a\Vert$ at $I_{(\vec x_0,r_0)}(\vec a)$ resp. 
$I'_{(\vec x_0,r_0)}(\vec a)$. 

The general strategy is then as follows: Take a set $S_i$, $i=1,\cdots N$,
of $N$ spheres with radii $r_i$ and centres $\vec x_i$, so that each 
sphere $S_i$ is disjoint from, and to the outside of, each other sphere
$S_j$, $j\ne i$. Take the constant function $f\equiv 1$ and take the sum over 
the free group generated by all $J_{(\vec x_i,r_i)}$ (alternatively 
the $J'_{(\vec x_i,r_i)}$). This converges to an analytic function $\phi$ 
provided $(N-1)r_*/d<1$, where $r_*=\mathrm{max}\{r_1,\cdots r_N\}$ and 
$d$ is the infimum of euclidean distances from the centres $\vec x_i$ to
points on the spheres $S_j$, $j\ne i$; 
see~\cite{Misner:1963}\cite{Giulini:1990}.\footnote{%
Note that $(N-1)r_*/d<1$ always holds in case of two spheres, $N=2$, 
since the condition of being disjoint and to the exterior of each other
implies $r_*/d<1$.} By construction the function $\phi$ is then invariant 
under each inversion map $J_{(\vec x_i,r_i)}$ (alternatively 
$J'_{(\vec x_i,r_i)}$). Consequently, the maps $I_{(\vec x_i,r_i)}$ 
(alternatively $I'_{(\vec x_i,r_i)}$) are isometries of the Metric 
$h:=\phi^4\,\delta$, which is defined on the manifold 
$\Sigma$ which one obtains by removing from $\reals^3$ the centres 
$\vec x_i$ and all their image points under the free group generated 
by the inversions $I_{(\vec x_i,r_i)}$ or $I'_{(\vec x_i,r_i)}$.
However, the topology of the manifold may be modified by suitable 
identifications using these isometries. For example, using  
$I'_{(\vec x_i,r_i)}$, we may excise the interiors of all spheres $S_i$
and identify antipodal points on each remaining boundary component 
$S_i$. In this fashion we obtain a manifold with one end, which is the 
connected sum of $N$ real-projective spaces minus a point (the point at 
spatial infinity). 

\begin{figure}
\centering\includegraphics[width=0.9\linewidth]{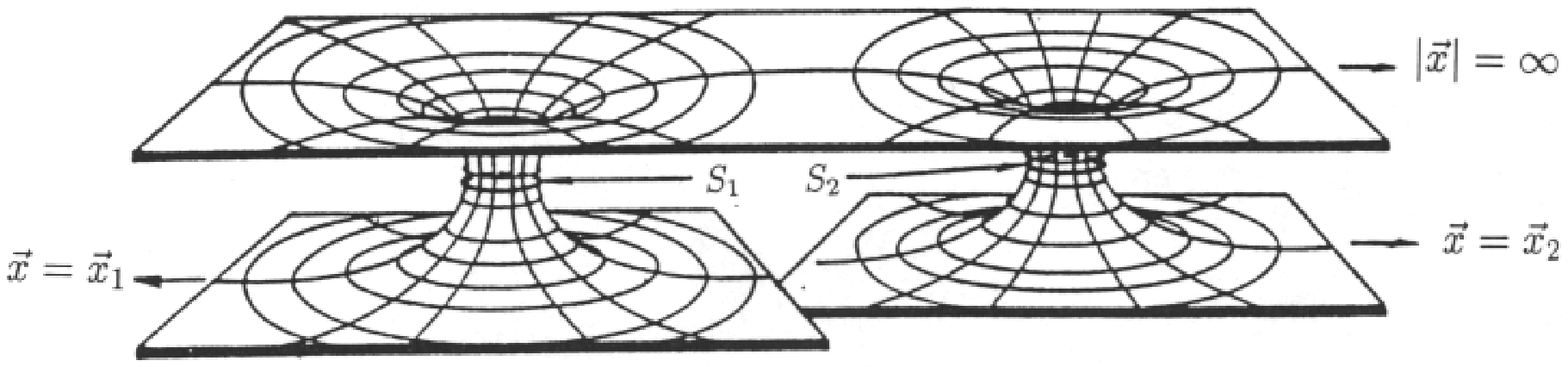}
\begin{minipage}[b]{0.49\linewidth}
\centering\includegraphics[width=1.0\linewidth]{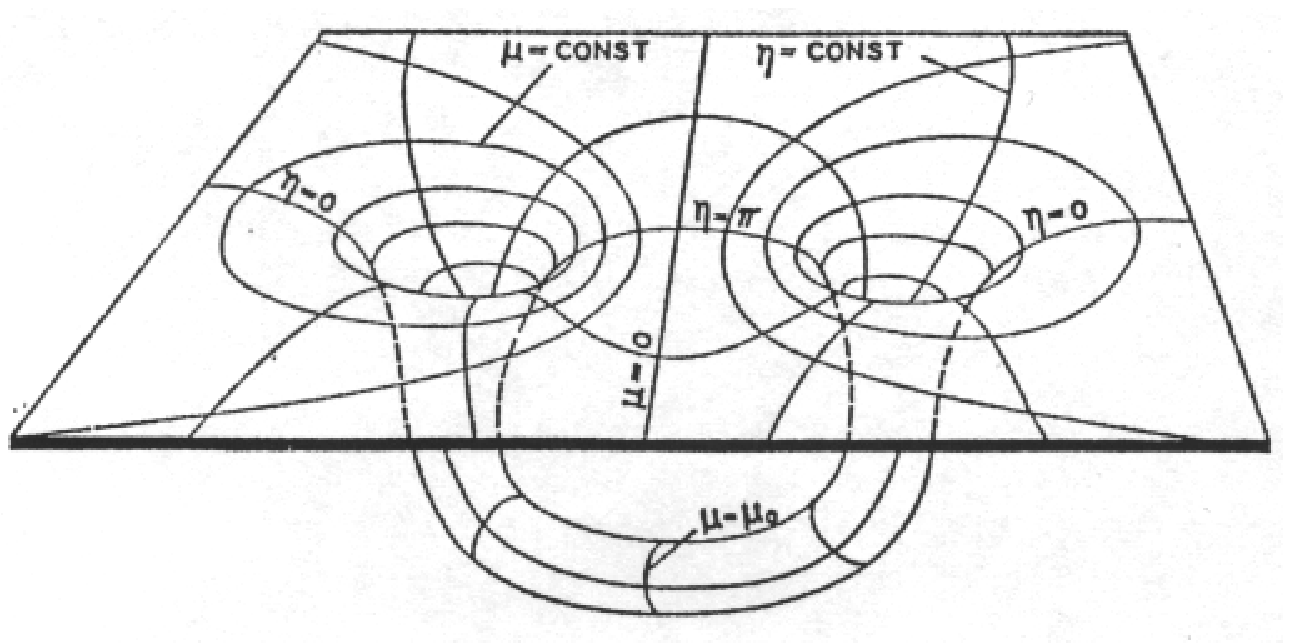}
\end{minipage}
\hfill
\begin{minipage}[b]{0.49\linewidth}
\centering\includegraphics[width=1.0\linewidth]{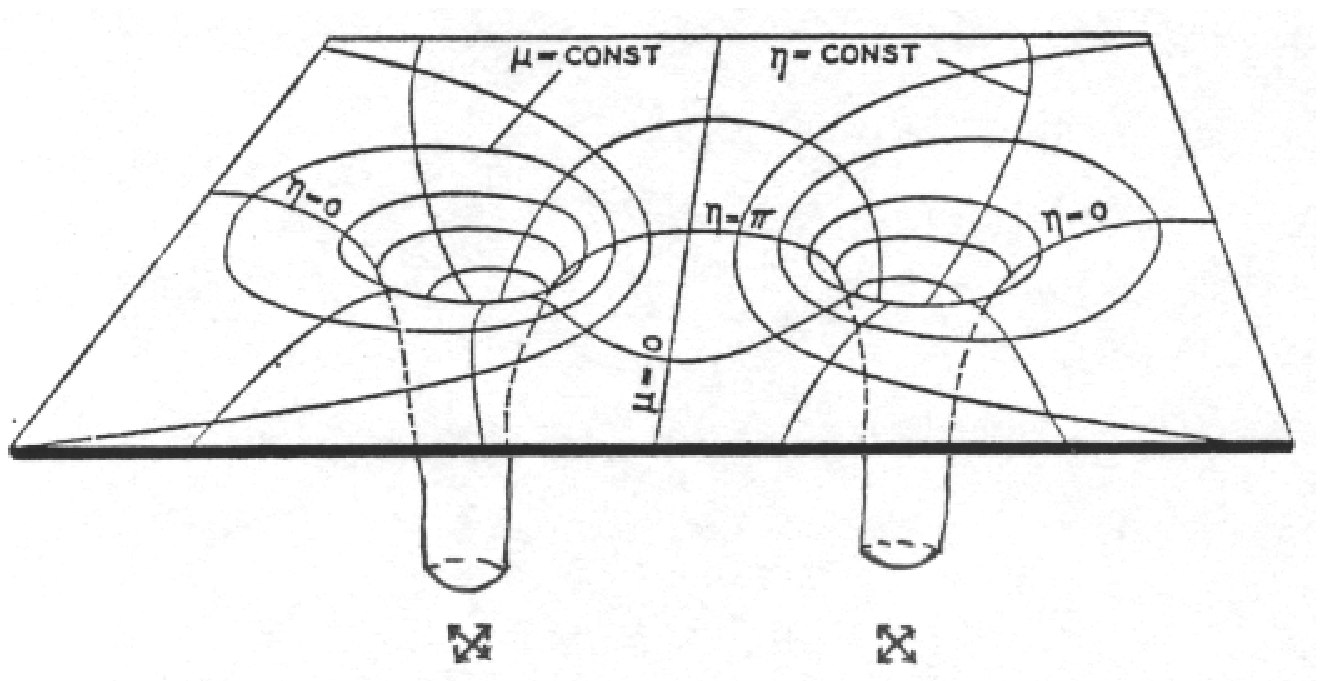}
\end{minipage}
\caption{\label{fig:TwoBlackHoles}%
Various topologies for data $(\Sigma,h)$ representing two black 
holes momentarily at rest. The upper manifold has three asymptotically
flat ends, one at spatial infinity and one each `inside' the apparent 
horizons (= minimal surfaces) $S_1,S_2$. The lower two manifolds have 
only one end each. The lower left manifold (wormhole) is topologically 
$S^1\times S^2-\{\mathrm{point}\}$ the lower right 
$\RP^3\#\RP^3-\{\mathrm{point}\}$, where $\#$ denotes connected sum.
The crosswise arrows in the lower right picture indicate that the 
shown 2-sphere boundaries are closed off by antipodal identifications.
The coordinates $\mu,\eta$ correspond to bispherical polar 
coordinates. No two of these three manifolds are locally isometric.}
\end{figure}

In general there are many topological options. Consider, for example, 
the simpler case of just two 2-spheres $S_1$ and $S_2$ of, say, 
equal radii, $r_1=r_2$. We again excise their interiors and identify 
their boundaries. If we use the maps $I_{(\vec x_i,r_i)}$ for the 
data construction, we may identify $S_1$ with $S_2$ in an orientation
\emph{reversing} fashion (with respect to their induced orientations) so 
that the quotient space is orientable. This results in Misner's 
wormhole\cite{Misner:1959} whose data are often used in numerical 
studies of black-hole collisions. If instead we had used the maps 
$I'_{(\vec x_i,r_i)}$ we have two choices: either to identify antipodal 
points on each $S_i$ separately, which results in the connected 
sum of two real-projective spaces, as explained above, or to 
identify $S_1$ with $S_2$, but now in an orientation preserving 
fashion (with respect to their induced orientations) so that the 
resulting manifold is a non-orientable version of Misner's wormhole 
discussed in~\cite{Giulini:1990}. The latter two manifolds are 
locally isometric but differ in their global topology, whereas 
they are not even locally isometric to the standard (orientable) 
Misner wormhole. 

Let us turn to the simplest non-trivial example: a single black hole. 
it corresponds to the solution of (\ref{eq:LaplaceEq}) with a single 
pole at, say, $\vec x_0=\vec 0$ and asymptotic value 
$\phi\rightarrow 1$ for $r\rightarrow\infty$, where $r:=\Vert\vec x\Vert$. 
Hence we have
\begin{equation}
\label{eq:SinglePolePhi}
\phi(\vec x)=1+\frac{m}{2r}\,.
\end{equation}
It is easy to verify that the constant $m$ just corresponds to the ADM 
mass defined via (\ref{eq:DefADMMass}). The 3-dimensional Riemannian 
manifold $(\Sigma,h)$ is now given by $\Sigma=\reals^3-\{\vec 0\}$ and 
the metric, in polar coordinates centred at the origin,
\begin{equation}
\label{eq:SinglePoleMetric}
h=\left(1+\frac{m}{2r}\right)^4\,\underbrace{
\bigl(dr^2 +r^2(d\theta^2+\sin^2\theta\,d\varphi^2)\bigr)}_{=\,\delta}
\end{equation}
It allows for the two discrete isometries
\begin{alignat}{2}
\label{eq:IsometryI}
&I:      \bigl(r,\theta,\varphi\bigr)
&&\mapsto\bigl(m^2/4r\,,\,\theta\,,\,\varphi\bigr)\,,\\
\label{eq:IsometryJ}
&J:      \bigl(r,\theta,\varphi\bigr)
&&\mapsto\bigl(m^2/4r\,,\,\pi-\theta\,,\,\varphi+\pi\bigr)\,.
\end{alignat}
The set of fixed points for $I$ is the sphere $r=m/2$, whereas $J$
acts freely (without fixed points). That the sphere $r=m/2$ is the 
fixed-point set of an isometry ($I$) implies that it is totally 
geodesic (has vanishing extrinsic curvature in $\Sigma$), as already 
discussed above. In particular it implies that $r=m/2$ is a minimal 
surface that joins two isometric halves. Hence $(\Sigma,h)$ has two 
asymptotically flat ends, one for $r\mapsto\infty$ (spatial infinity)
and one for $r\mapsto 0$, as shown on the left of 
Fig.\,\ref{fig:Kruskal}. This is sometimes interpreted by saying 
that there is a singular pointlike mass source at $r=0$, just like for 
the electric Coulomb field for a point charge. But this interpretation is 
deceptive. It is true that the Coulomb field is a vacuum solution to 
Maxwell's equations if the point at which the source sits is simply 
removed from space. But this removal of a point leaves a clear trace 
in that the resulting manifold is incomplete. This is different for 
the manifold $(\reals-\{\vec 0\}\,,\,h)$, with $h$ given by  
(\ref{eq:SinglePoleMetric}), which is complete, due to the fact that 
the origin is infinitely far away in the metric $h$. Hence no point is 
missing and the solution can be regarded as a genuine vacuum solution. 

\subsection{The  $\RP^3$ geon}
\label{sec:RP3Geon}
\begin{figure}[ht]
\begin{minipage}[c]{0.48\linewidth}
\centering\includegraphics[width=1.0\linewidth]{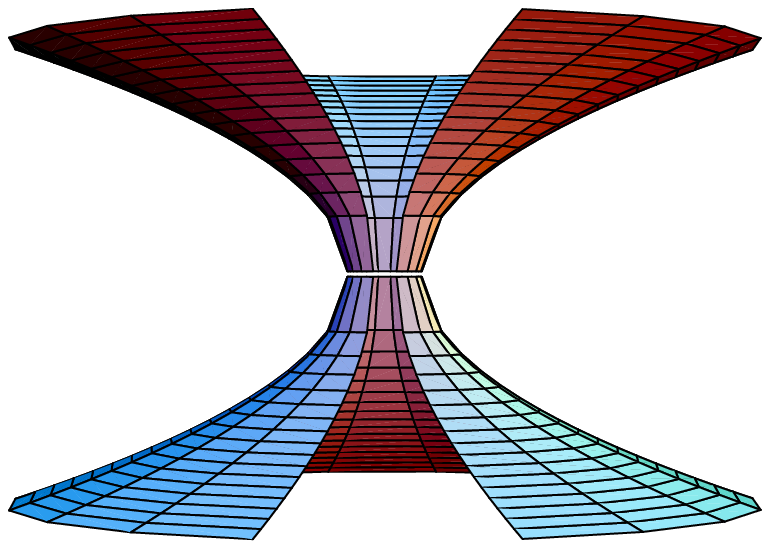}
\end{minipage}
\hfill
\begin{minipage}[c]{0.50\linewidth}
\centering
\includegraphics[width=1.0\linewidth]{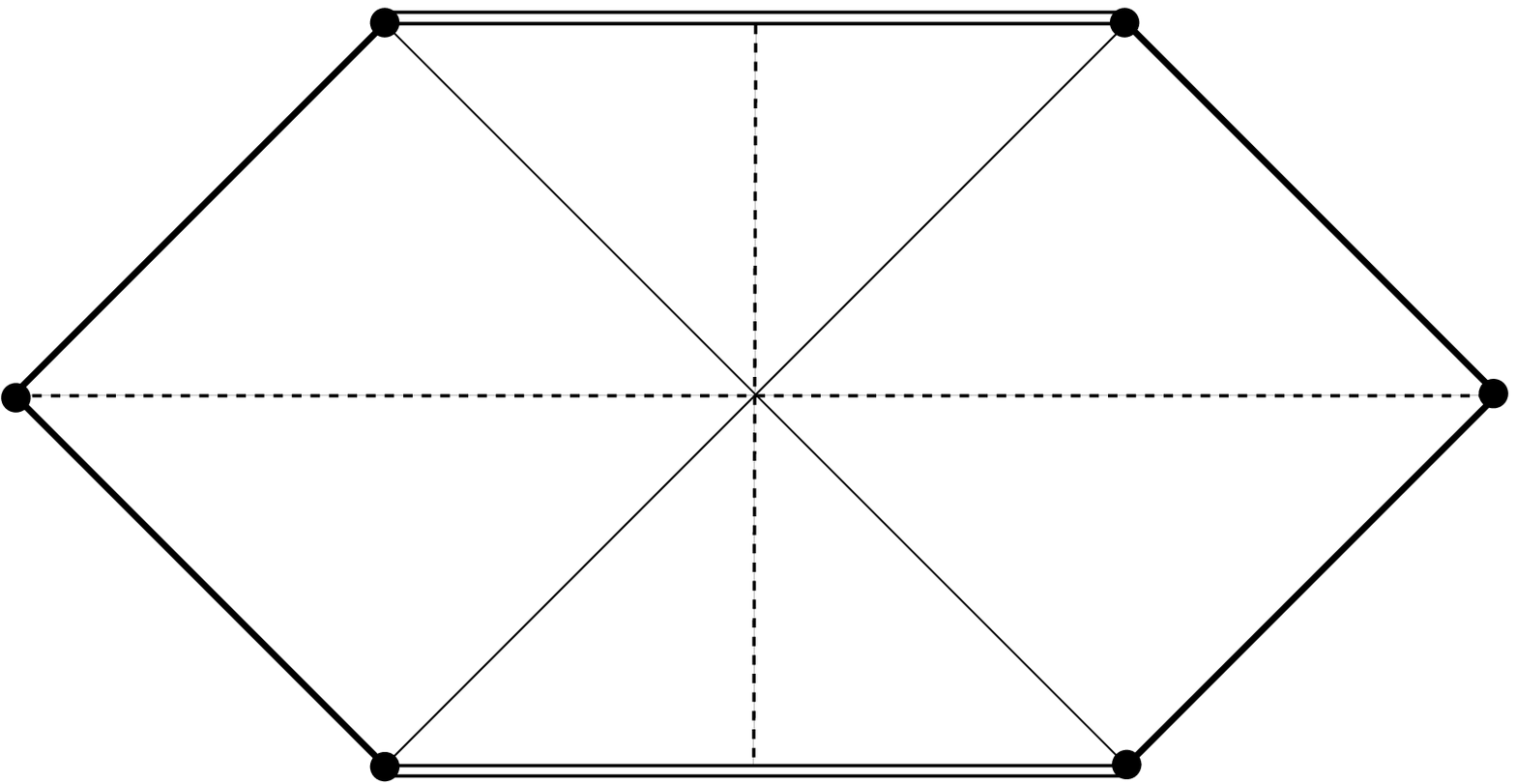}
\end{minipage}
\put(-135,4){\tiny $T=0$}
\put(-86,18){\tiny\begin{rotate}{90}$X=0$\end{rotate}}
\put(1,2){\tiny $i_0$}
\put(-175,2){\tiny $i_0$}
\put(-47,51){\tiny $i_+$}
\put(-129,51){\tiny $i_+$}
\put(-47,-48){\tiny $i_-$}
\put(-129,-48){\tiny $i_-$}
\put(-20,23){\tiny $I^+$}
\put(-20,-21){\tiny $I^-$}
\put(-155,23){\tiny $I^+$}
\put(-157,-21){\tiny $I^-$}
\put(-45,15){$\mathrm{I}$}
\put(-75,30){$\mathrm{II}$}
\put(-132,15){$\mathrm{III}$}
\put(-75,-30){$\mathrm{IV}$}
\put(-275,40){$T=0$}
\put(-305,1){\tiny $X=0\quad\rightarrow$}
\caption{\label{fig:Kruskal}%
To the right is the conformal (Penrose) diagram of Kruskal spacetime 
in which each point of this 2-dimensional representation corresponds 
to a 2-sphere (an orbit of the symmetry group of spatial rotations). 
The asymptotic regions are $i_0$ (spacelike infinity), $I^{\pm}$ 
(future/past lightlike infinity), and $i^{\pm}$ (future/past timelike 
infinity). The diamond and triangular shaped regions $\mathrm{I}$
and $\mathrm{II}$ correspond to the exterior ($r>2m$) and interior 
($0<r<2m$) Schwarzschild spacetime respectively, the interior being 
the black hole. The triangular region $\mathrm{IV}$ is the time reverse of 
$\mathrm{II}$, a white hole. Region $\mathrm{III}$ is another 
asymptotically flat end isometric to the exterior Schwarzschild region
$\mathrm{I}$. The double horizontal lines on top an bottom represent the 
singularities ($r=0$) of the black and white hole respectively. 
The left picture shows an embedding diagram of the hypersurface $T=0$ 
(central horizontal line in the conformal diagram) that serves to 
visualise its geometry. Its minimal 2-sphere at the throat corresponds 
to the intersection of the hyperplanes $T=0$ and $X=0$ (bifurcate 
Killing Horizon).}
\end{figure}

There is a different twist to this story. One might object against 
the fact that $\reals-\{\vec 0\}$ has \emph{two} ends rather than
just one (at spatial infinity). After all, what would the `inner 
end' correspond to? A locally isometric manifold with just one
end is obtained by taking the quotient of  $\reals-\{\vec 0\}$
with respect to the freely acting group $\integers_2$ that is 
generated by the isometry $J$ in (\ref{eq:IsometryJ}). This identifies 
the region $r>m/2$ with the region $r<m/2$ and antipodal points on the 
minimal 2-sphere $r>m/2$. The resulting space is real projective 
3-space, $\RP^3$,  minus a point, which clearly has 
just one end. Its full time evolution, i.e. the space-time emerging 
from it, can be obtained from the maximal evolution of the Schwarzschild 
data: $h$ as in (\ref{eq:SinglePoleMetric}),  $K=0$, which is Kruskal 
spacetime (see \cite{Kruskal:1960} and/or Chapter\,5.5. in 
\cite{Hawking.Ellis:TLSSOS}). A conformally rescaled version 
(Penrose Diagram) of Kruskal spacetime is depicted on the right 
of Fig.\,\ref{fig:Kruskal}.

In Kruskal coordinates\footnote{Kruskal~\cite{Kruskal:1960} 
uses $(v,u)$ Hawking Ellis \cite{Hawking.Ellis:TLSSOS} $(t',x')$
for what we call $(T,X)$.} $(T,X,\theta,\varphi)$, where $T$ and $X$ 
each range in $(-\infty,\infty)$ obeying $T^2-X^2<1$, the Kruskal 
metric reads (as usual, we write 
$d\Omega^2$ for $d\theta^2+\sin^2\theta\,d\varphi^2$): 
\begin{equation}
\label{eq:KruskalMetric1}
g=\frac{32 m^2}{\rho}\,\exp(-\rho/2m)\,\bigl(-dT^2+dX^2\bigr)+r^2d\Omega^2\,,
\end{equation}
where $\rho$ is a function of $T$ and $X$, implicitly 
defined by 
\begin{equation}
\label{eq:KruskalMetric2}
\bigl((\rho/2m)-1\bigr)\,\exp(\rho/2m)=X^2-T^2\,.
\end{equation}
Here $\rho$ corresponds to the usual radial coordinate, in terms of 
which the Schwarzschild metric reads
\begin{equation}
\label{eq:SchwCoordinatesMetric}
g=-\left(1-\frac{2m}{\rho}\right)\,dt^2
+\frac{dr^2}{1-\frac{2m}{\rho}}+\rho^2\,d\Omega^2\,
\end{equation}
where $\rho>2m$. It covers region~I of the Kruskal spacetime.
Setting 
\begin{equation}
\label{eq:IsotropRadialCoord}
\rho=r\left(1+\frac{m}{2r}\right)^2
\end{equation}
so that the range $m/2<r<\infty$ covers the range $2m<\rho<\infty$
twice, we obtain the `isotropic form'
\begin{equation}
\label{eq:IsoSchwCoordinatesMetric}
g=-\left(\frac{1-\frac{m}{2r}}{1+\frac{m}{2r}}\right)^2\,dt^2
+\left(1+\frac{m}{2r}\right)^4\bigl(dr^2+r^2\,d\Omega^2\bigr)
\end{equation}
which covers regions I and III of the Kruskal manifold. 

The Kruskal metric (\ref{eq:KruskalMetric1}) is spherically 
symmetric and allows for the additional Killing field%
\footnote{That $K$ is Killing is immediate, 
since~(\ref{eq:KruskalMetric2}) shows that $\rho$ depends 
only on the combination $X^2-T^2$ which is clearly annihilated by 
$K$.} 
\begin{equation}
\label{eq:KruskalKilling}
K=\tfrac{1}{4m} \bigl(X\partial_T+T\partial_X\bigr)\,,
\end{equation}
which is timelike for $\vert X\vert>\vert T\vert$ and spacelike for 
$\vert X\vert<\vert T\vert$.

The maximal time development of the $\RP^3$ initial data set is 
now obtained by making the following identification on the Kruskal
manifold:

\begin{equation}
\label{eq:KrskalZ2Isometry}
J:(T,X,\theta,\varphi)\mapsto (T,-X,\pi-\theta,\varphi+\pi)\,.
\end{equation}
It generates a freely acting group $\mathbb{Z}_2$ of smooth 
isometries which preserve space- as well as time-orientation. 
Hence the quotient is a smooth space- and time-orientable manifold, 
the $\RP^3$\emph{geon}.\footnote{The $\RP^3$ geon is different 
from the two mutually different `elliptic interpretations' of the 
Kruskal spacetime discussed in the literature by Rindler, Gibbons, 
and others. In \cite{Rindler:1965} the identification map considered 
is $J':(T,X,\theta,\varphi)\mapsto(-T,-X,\theta,\varphi)$, which 
gives rise to singularities on the set of fixed-points 
(a 2-sphere) $T=X=0$. Gibbons \cite{Gibbons:1986} takes  
$J'':(T,X,\theta,\varphi)\mapsto(-T,-X,\pi-\theta,\varphi+\pi)$,
which is fixed-point free, preserves the Killing field 
(\ref{eq:KruskalKilling}) (which our map $J$ does not), but does 
not preserve time-orientation. $J''$ was already considered in 
1957 by Misner \& Wheeler (Section\,4.2 in \cite{Misner.Wheeler:1957}), 
albeit in isotropic Schwarzschild coordinates already mentioned above, 
which only cover the exterior regions $\mathrm{I}$ and $\mathrm{III}$ 
of the Kruskal manifold.}
Its conformal diagram is just given by 
cutting away the $X<0$ part (everything to the left of the vertical 
$X=0$ line) in Fig.\,\ref{fig:Kruskal} and taking into account that 
each point on the remaining edge, $X=0$, now corresponds to a 2-sphere
with antipodal identification, i.e. a $\RP^2$ (which is not orientable). 
The spacelike hypersurface $T=0$ has now the topology of the 
once punctured $\RP^3$. In the left picture of Fig.\,\ref{fig:Kruskal}
this corresponds to cutting away the lower half and eliminating the 
inner boundary 2-sphere $X=0$ by identifying antipodal points. 
The latter then becomes a minimal one-sided non-orientable surface 
in the orientable space-section of topology $\RP^3-\{\mathrm{point}\}$.
The $\RP^3$ geon isometrically contains the exterior Schwarzschild 
spacetime (region\,$\mathrm{I}$) with timelike Killing field $K$.
But $K$ ceases to exits globally on the geon spacetime since it 
reverses direction under (\ref{eq:KrskalZ2Isometry}). 

\section{$X$ without $X$}
\label{sec:XwithoutX}

\subsection{Mass without mass}
\label{sec:MassWithoutMass}
At the end of Section~\ref{sec:VacuumData} we already 
explained in what sense (active gravitational) mass emerges 
from pure topology and the constraints implied by Einstein's 
equation. Physically this just means that localised 
configurations of overall non-vanishing mass/energy may be 
formed from the gravitational field alone. With some care one 
may say that such solutions represent bounded states of gravitons
(`graviton balls'). However, they cannot be stable since 
Gravitational solitons do not exist (in four spacetime 
dimensions)!  

If $\Sigma$ is topologically non-trivial, Gannon's 
theorem~\cite{Gannon:1975} (already discussed above) implies 
in full generality that the spacetime is singular 
(geodesically incomplete). The non-existence of vacuum, 
stationary, asymptotically flat spacetimes with non-vanishing 
mass, where the spacetime topology is $\reals\times\Sigma$ 
and where $\Sigma$ has only one end (spatial infinity), follows 
immediately from the expression (\ref{eq:DefKomarMass}) for the 
overall mass. Indeed, converting the surface integral 
(\ref{eq:DefKomarMass}) into a space integral via Stokes' theorem 
and using that $d\star dK^\flat$ is proportional to the 
spacetime's Ricci tensor shows%
\footnote{One uses the Killing identity 
$\nabla_a\nabla_b K_c=K_dR^d_{\phantom{d}a\,bc}$ to convert
the second derivatives of $K^\flat$ into terms involving no 
derivatives and the Riemann tensor.} 
that the expression vanishes identically due to the 
source-free Einstein equation. This generalises an older 
result due to Einstein \& Pauli~\cite{Einstein.Pauli:1943}
and is known as `Lichnerowicz theorem', since Lichnerowicz 
first generalised the Einstein \& Pauli result from static
to stationary spacetimes, albeit using a far more involved
argument than that given here 
(see~\cite{Lichnerowicz:TheoriesRelativistes1955}, 
livre premier, chapitre~VIII). 

Most interestingly, this non-existence result ceases to be 
true in higher dimensions, as is exemplified by the existence 
of so-called Kaluza-Klein monopoles~\cite{Sorkin:1983,Gross.Perry:1983}, 
which are non-trivial, regular, static, and `asymptotically flat' 
solutions to the source-free Einstein equations in a five-dimensional 
spacetime. The crucial point to be observed here is that 
the Kaluza-Klein spacetime is `asymptotically flat' in the 
sense that it is asymptotically flat in the ordinary sense 
for three spatial directions, but not in the added fourth 
spatial direction, which is topologically a circle.
Had asymptotic flatness in $n$ dimensional spacetime been required 
for all $n-1$ spatial directions, no such solution could 
exist~\cite{Deser:1988}.

In this connection it is interesting to note that in their 
paper~\cite{Einstein.Pauli:1943}, Einstein \& Pauli actually 
claim to show the non-existence of soliton-like solutions in 
all higher dimensional Kaluza-Klein theories even though they 
require asymptotic flatness in three spatial directions. 
But closer inspection reveals that their proof, albeit correct, 
invokes an additional and physically unjustified topological 
hypothesis that is violated by Kaluza-Klein monopoles. This is 
explained in more detail in~\cite{Giulini:2008a}. Hence we may 
take Kaluza-Klein monopoles as a good example for the generation 
of mass and also magnetic charge in the framework of pure (higher
dimensional!) General Relativity without any sources.

\subsection{Momenta without momenta}
\label{sec:MomentaWithoutMomenta}
Source free solutions with linear and angular momenta are also
not difficult to obtain. Let us here just note a simple way 
of how to arrive at flux-integral expressions for these 
quantities. Let again $(h,\pi)$ be a data set which is asymptotically 
flat on $\Sigma$ with one end. Let $\xi$ be a vector field on 
$\Sigma$ that tends to the generator of an asymptotic isometry 
at infinity, that is, either a translation or a rotation. 
The corresponding linear or angular momentum is then just given 
by the usual \emph{momentum map} corresponding to $\xi$:
\begin{equation}
\label{eq:MomentumMap1}
\xi\mapsto \int_\Sigma d^3x\,\pi^{ab}\,L_\xi h_{ab}=:p_\xi\,,
\end{equation}
where the right hand side is considered as a function on phase space
$T^*\Riem$. Using the momentum constraint, $\nabla_a\pi^{ab}=0$,
an integration by parts in (\ref{eq:MomentumMap1}) converts it into a 
flux integral at spatial infinity which, re-expressing $\pi$ in 
terms of $K$, reads
\begin{equation}
\label{eq:MomentumMap2}
p_\xi:=\lim_{R\rightarrow\infty}
\left\{
\frac{1}{8\pi}\int_{S_R}\bigl(
K_{ij}-h_{ij}h^{ab}K_{ab}\bigr)\xi^in^j\,d\Omega
\right\}\,.
\end{equation}
This is the well known expression for the ADM (Arnowitt, Deser, Misner)
linear and angular momentum in geometric 
units~\footnote{That is, linear momentum has the unit of length 
(like mass) and angular momentum of length-squared. The are converted 
into ordinary units through multiplication with $c/G$.}. 

Obviously there cannot exist a non-trivial asymptotically flat initial data
set with an exact translational symmetry (because that translation 
could shift any local lump of curvature arbitrarily far into the 
asymptotically flat region, so that the curvature must be zero).
But there may be such data sets with exact rotational symmetry. 
In that case, if $\xi$ is the rotational Killing field and 
$\xi^\flat:=g(\xi,\,\cdot\,)$ its associated one form in spacetime,  
a much simpler expression for angular momentum is given by the 
Komar integral~\cite{Komar:1959}:
\begin{equation}
\label{eq:KomarIntegral2}
p_\xi:=\lim_{R\rightarrow\infty}
\left\{
\frac{1}{16\pi}\int_{S_R}\star\, d\xi^\flat\right\}\,,
\end{equation}
where, as before, $d$ is the exterior differential in spacetime and $\star$ 
the Hodge dual with respect to the spacetime metric $g$. Again 
$d\star d\xi^\flat$ is proportional to the spacetime's Ricci tensor 
and hence zero, since spacetime is assumed to satisfy the source
free Einstein equation. Hence Stokes' theorem implies that if $\Sigma$
has only one end (spatial infinity) and the solution is regular 
in the interior, $p_\xi$ must be zero. Therefore there cannot exist 
regular data set which give rise to rotationally symmetric solutions 
with non-vanishing angular momenta. A minimal relaxation is given by 
data sets which are \emph{locally} symmetric, that is, in which a rotational 
Killing field exists up to sign. This slight topological generalisation
indeed suffices to render the non-existence argument just given 
insufficient. Such data sets with net angular momentum have been 
constructed in~\cite{Friedman.Mayer:1982}. 

\subsection{Charge without charge}
\label{sec:ChargeWithoutCharge}
One case of `charge without charge' is clearly given by the Kaluza-Klein 
monopoles mentioned above. Here we wish to stick to four spacetime
dimensions and ask whether electric or magnetic charge can emerge from 
the Einstein-Maxwell equations without sources for the Maxwell field
(in distinction to above, the energy-momentum tensor for the Maxwell
field now acts as a source for the gravitational field). 

If $F$ is the 2-form on spacetime that represents the electromagnetic 
field, then the electric and magnetic charges $q_e$ and $q_m$ inside
a 2-sphere $S$ are respectively given by
\begin{subequations}   
\label{eq:EM_Charges}
\begin{alignat}{2}
\label{eq:EM_Charges-a}
& q_e &&= \frac{1}{4\pi}\int_S \star F\,,\\
\label{eq:EM_Charges-b}
& q_m &&= \frac{1}{4\pi}\int_S F\,.
\end{alignat}
\end{subequations}
Since $dF=0$ (homogeneous Maxwell equation) and $d\star F=0$ 
(inhomogeneous Maxwell equation with vanishing sources) these integrals
depend only on the homology class on $S$. This seems to imply 
that if spacetime has a regular interior, i.e. is of topology 
$\reals\times\Sigma$ and $\Sigma$ has only one end 
(spatial infinity), there will be no global net charge. The 
only possibility seems to be that there are local charges, like, 
e.g., if $\Sigma$ has a wormhole topology 
$S^1\times S^2-\{\mathrm{point}\}$, as shown by the lower-left 
drawing in Fig.\,\ref{fig:TwoBlackHoles}, where the flux lines thread 
through the wormhole. The homology class of 2-spheres that contain 
both wormhole mouths has zero charge, whereas the two individual 
wormhole mouths have equal and opposite charges associated to 
them. 

However, this is not the only possibility! Our argument above 
relied on Stokes' theorem, which for ordinary forms presumes that 
the underlying manifold is orientable. On a non-orientable 
manifold it only holds true for forms of density weight one, i.e. 
sections of the tensor bundle of forms twisted with the (now non 
trivial) orientation bundle; 
see, e.g., \S\,7 of~\cite{Bott.Tu:DiffFormsAlgTop}. 

\begin{figure}[htb]

\centering\includegraphics[width=0.75\linewidth]{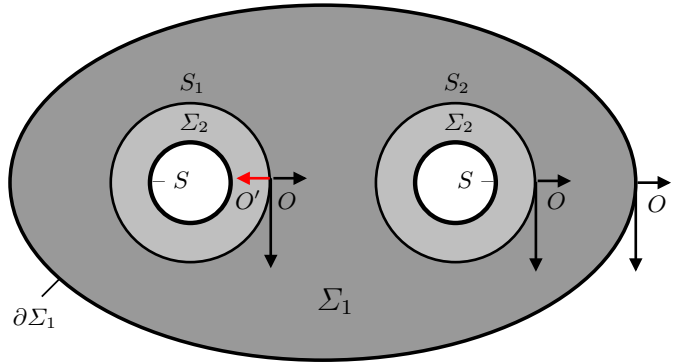}
\put(-82,66){$S$ --}
\put(-197,66){-- $S$}
\put(-187,103){$S_1$}
\put(-87,103){$S_2$}
\put(-90,88){ $\Sigma_2$}
\put(-187,88){$\Sigma_2$}
\put(-135,20){\large $\Sigma_1$}
\put(-10,57){$O$}
\put(-48,58){$O$}
\put(-150,58){$O$}
\put(-166,58){$O'$}
\put(-250,14){$\partial\Sigma_1$}
\caption{\label{fig:NonOrientableWormehole} %
Consider the three-dimensional region that one obtains
by rotating this figure about the central horizontal 
axis of symmetry. The two inner boundary spheres $S$ are to be identified
in a way so that their induced orientations $O$ match, e.g. by
simple translation. (In this two-dimensional picture an
orientation is represented by an ordered two-leg, where the
ordering is according to the different lengths of the legs.)
This results in a \emph{non-orientable} 
manifold with single outer boundary component 
$\partial\Sigma_1$, corresponding to the non orientable 
wormhole. In the text we apply Stokes' theorem twice to two orientable 
submanifolds: First, to the heavier shaded region bounded by
the outer 2-sphere $\partial\Sigma_1$ with orientation $O$
and the inner two 2-spheres $S_1$ and $S_2$ with like 
orientations $O$ as indicated. Second, to the lightly shaded 
cylindrical region labelled by $\Sigma_2$ that is bounded by 
the two 2-spheres $S_1$ and $S_2$ with opposite 
orientations $O'$ and $O$ respectively.}
\end{figure}

This means that the argument for the non-existence of global 
charges can be extended to the non-orientable case for $\star F$
(which is a two form of density weight one) but not to $F$
(which is a two form of density weight zero). Hence net 
electric charges cannot, but magnetic\footnote{The distinction
between electric and magnetic is conventional in Einstein-Maxwell
theory without sources for the Maxwell field, since the energy-momentum 
tensor $T$ for the latter is invariant under duality 
rotations which rotate between $F$ and $\star F$ according to 
$\omega\mapsto\exp(i\varphi)\,\omega$, where 
$\omega:=F+i\,\star F$. Since $T_{\mu\nu}\propto
\omega_{\mu\lambda}\bar\omega_\nu^{\phantom{\nu}\lambda}$,
where an overbar denotes complex conjugation, the invariance of $T$
is immediate.} can exist~\cite{Sorkin:1977}\cite{Friedman.Mayer:1982}. 
A simple illustration of how orientability comes into this 
is given by Fig.\,\ref{fig:NonOrientableWormehole}.
(For the possibility to have net electric charge due to 
non \emph{time}-orientable spacetime manifolds 
compare~\cite{Diemer.Hadley:1999}.)

As stated above, in the non-orientable case, 
Stokes theorem (here in three dimensions) continues to apply 
to two-forms of density weight one (e.g. the Hodge duals of one 
forms) and does not apply to two-forms of density weight 
zero, like the magnetic two form or, equivalently, its 
Hodge dual, which is a pseudo-vector field $\vec B$ of zero 
divergence. We apply Stokes' theorem to suitable 
orientable submanifolds as explained in the caption to 
Fig.\,\ref{fig:NonOrientableWormehole}. We obtain,
denoting the flux of $\vec B$ through a surface $S$
with orientation $O$ by $\Phi(\vec B,S,O)$,
\begin{equation}
\label{eq:FluxIntegral1} 
\Phi(\vec B,\partial\Sigma_1,O)+
 \Phi(\vec B,S_1,O)+ \Phi(\vec B,S_2,O)=0
\end{equation}
in the first case, and 
\begin{equation}
\label{eq:FluxIntegral2} 
 \Phi(\vec B,S_1,O')+ \Phi(\vec B,S_2,O)=0
\end{equation}
in the second. Using the obvious fact that the flux integral 
changes sign if the orientation is reversed, i.e., 
$\Phi(\vec B,S_1,O')=-\Phi(\vec B,S_1,O)$, we get 

\begin{equation}
\label{eq:FluxIntegral3} 
 \Phi(\vec B,\partial\Sigma_1,O)=-2\,\Phi(\vec B,S_1,O)\,.
\end{equation}
So in order to get a non-zero global charge, we just 
need to find a divergenceless pseudo-vector field on 
$\Sigma_1$ with non vanishing flux through $S$, which 
can be arranged.  

Note that the trick played here in using non orientable 
$\Sigma$ would not work for the Komar integrals (\ref{eq:DefKomarMass})
(\ref{eq:KomarIntegral2}), since the Hodge map $\star$ turns the 
ordinary two form $dK^\flat$ (or $d\xi^\flat$) of density weight 
zero into the two form $\star dK^\flat$ (or $\star d\xi^\flat$) of 
density weight one, so that Stokes' theorem continues to hold
in these cases for for non-orientable $\Sigma$ by the result 
cited above. 

\subsection{Spin without spin}
\label{sec:SpinWithoutSpin}
In my opinion the by far most surprising  case of `$X without X$'
if that where $X$ stands for \emph{spin}, i.e., half-integral angular 
momentum. It was certainly not anticipated by Misner, Thorne, and 
Wheeler, who in their otherwise most comprehensive 
book~\cite{Misner.Thorne.Wheeler:Gravitation} were quite lost 
in trying to answer their own question of how ``to find a natural
place for spin\,1/2 in Einstein's standard geometrodynamics
(Box\,44.3 in \cite{Misner.Thorne.Wheeler:Gravitation}). 
A surprising answer was offered 8 years later, in 1980, by 
John Friedman and Rafael Sorkin~\cite{Friedman.Sorkin:1980}. 

It is often said that the need to go from the group $SO(3)$ of spatial
rotations to its double (= universal) cover, $SU(2)$, is 
quantum-mechanical in origin and cannot be understood on a classical 
basis. In some sense the mathematical facts underlying the idea of 
`spin\,1/2 from gravity' disprove this statement. They imply that
if the 3-manifold $\Sigma$ has a certain topological characteristic, 
the asymptotic symmetry group for isolated systems (modelled by spatially 
asymptotically flat data) is not the Poincar\'e group (inhomogeneous
Lorentz group) in the sense of~\cite{Beig.Murchadha:1987}, but 
rather its double (= universal) covers -- for purely topological 
reasons! Let us try to explain all this in more detail. 

Recall that in Quantum Mechanics the possibility for this 
enlargement (central extension) of a classical symmetry group has 
its origin in the assumption that the phase of the complex wave 
function is a redundant piece of description (i.e. unobservable), 
at least for states describing isolated systems, so that symmetry 
groups should merely act on the space of rays rather than on Hilbert space 
by proper representations. Hence it is sufficient for the  
symmetry group to be implemented by so-called ray 
representations, which in case of the rotation group are in 
bijective correspondence to proper representations of its 
double (= universal) cover group. Accordingly, in Quantum 
Mechanics, there exist physically relevant systems whose state 
spaces support proper representations of $SU(2)$ but not of 
$SO(3)$: These are just the systems whose angular momentum is 
an odd multiple of $\hbar/2$. We will say that such systems 
admit \emph{spinorial states}. 

Spinorial states are not necessarily tight to the usage of spinors. 
They also have a place in ordinary Schr\"odinger quantisation, i.e. 
for systems whose quantum state space is represented by the Hilbert 
space of square integrable functions over the classical configuration 
space $Q$, which here and in what follows is understood to be the 
reduced configuration space in case constraints existed initially. 
Then, spinorial states exist if the following conditions 
hold: 
\begin{itemize}
\item[S1]
$Q$ is not simply connected.
\item[S2]
The (say left) action $SO(3)\times Q\rightarrow Q$, 
$(g,q)\mapsto g\cdot q$, of the ordinary rotation group $SO(3)$ on 
the classical configuration space $Q$ is such that if 
$\gamma: [0,2\pi]\mapsto SO(3)$ is any full 360-degree rotation about 
some axis, then the loop $\Gamma:=\gamma\cdot q$ in $Q$, based at $q\in Q$,
is not contractible, i.e. defines a non-trivial element (of order two
since $\gamma$ traversed twice is contractible in $SO(3)$) in 
$\pi_1(Q,q)$, the fundamental group of $Q$ based at $q$. It is not
hard to see that this property (of being non-contractible) is 
independent of the basepoint $q\in Q$ within the same path component of
$Q$, though it may vary if one goes from one path component to another
(as in the Skyrme model mentioned below). See \cite{Giulini:1993}, 
in particular the proof of Lemma\,1.  
\end{itemize}
 
The reason for the existence of spinorial states in such 
situations lies in possible generalisations of Schr\"odinger
quantisation if the domain for the wave function is a 
space, $Q$, whose fundamental group is non trivial. The idea of 
generalisation is to define the Schr\"odinger function on the 
universal cover space $\bar Q$ (i.e. the Hilbert space is the 
space of square-integrable functions on $\bar Q$) but 
to restrict the observables to those that commute with the 
unitary action of the deck transformations. The latter then 
form a discrete gauge group isomorphic to the fundamental 
group of $Q$. The Hilbert space decomposes into superselection
sectors which are labelled by the equivalence classes of unitary
irreducible representations. The sector labelled by the trivial 
class is isomorphic to that of ordinary Schr\"odinger quantisation
on $Q$, whereas the other sectors are acquired through the 
generalisation discussed here.

This is related to, but not identical with, another generalisation 
that is usually mentioned in the context of geometric quantisation. 
There one generalises Schr\"odinger quantisation by considering 
quantum states as square-integrable \emph{sections} in a complex line 
bundle over $Q$ (rather than just complex-valued functions on $Q$). 
This leads to additional sectors labelled by the equivalence classes 
of complex line bundles, which are classified by $H^2(Q,\integers)$,
the second cohomology group of $Q$ with integer coefficients 
(see, e.g., \cite{Woodhouse:GQ}). 

In generalised Schr\"odinger quantisation spinorial states will 
correspond to particular such new sectors. To make this more precise 
in the geometric-quantisation picture, we recall that $H^2(Q,\integers)$, 
being a finitely generated abelian group, has the structure 
\begin{equation}
\label{eq:StructureH2}
H^2(Q,\integers)\cong
\underbrace{\integers\oplus\cdots\oplus\integers}_{\text{free part}}\,\oplus\,
\underbrace{\integers_{p_1}\oplus\cdots\integers_{p_n}}_{\text{torsion part}}\,.
\end{equation}
The number of factors $\integers$ in the free part is called the second 
Betti number and the number $n$ of cyclic groups the second torsion 
number. For this to be well defined we have to agree that each of the 
integers $p_i$ should be a power of a prime.%
\footnote{A classic theorem on finite abelian groups states that if 
$p,q$ are integers, $\integers_{pq}$ is isomorphic to 
$\integers_p\oplus\integers_q$ iff $p$ and $q$ are coprime.} 
Spinorial states are then given by sections in all those line bundles 
which represent \emph{a particular} $\integers_2$ factor in the torsion 
part of its decomposition according to (\ref{eq:StructureH2}) non 
trivially.

We said `a particular $\integers_2$ factor'. Which one? The answer is:
That one, which is generated by the 360-degree rotation according to 
criterion S2 above. To understand this, we remark that the torsion part 
of $H^2(Q,\integers)$ can be understood in terms of the fundamental 
group. More precisely, the torsion part of $H^2(Q,\integers)$ is 
isomorphic to the torsion part of the abelianisation of the 
fundamental group.%
\footnote{This follows in two steps: First one recalls 
$H^2(Q,\integers)$ is isomorphic to the direct sum of the 
free part of $H_2(Q,\integers)$ and the torsion part of 
$H_1(Q,\integers)$ (universal coefficient theorem). Second
one uses that $H_1(Q,\integers)$ is isomorphic to the 
abelianisation of the fundamental group (Hurewicz' theorem).}
Given that isomorphism, we can now identify the $\integers_2$
factor in $H^2(Q,\integers)$ with the $\integers_2$ subgroup of
the fundamental group that is generated by 360-degree rotations,
as explained by S2. 

A simple illustrative example of this is given by the rigid rotor, that 
is a system whose configuration space $Q$ is the group manifold 
$SO(3)$, which as manifold is isomorphic to $\RP^3$. The action 
of physical rotations is then given by left translation. Here we 
have  $H^2(Q,\integers)\cong\integers_2$, i.e. it is pure torsion 
and, in fact, isomorphic to the fundamental group. Quantisation 
then leads to two sectors: Those containing states of integral spin, 
which are represented by ordinary square integrable functions on $Q$,
and those containing half-integral spin, represented by square 
integrable sections in the unique non-trivial line bundle over
$Q\cong\RP^3$. 

More sophisticated field theoretic examples for this mechanism 
are given by so-called 
non-linear sigma models, in which the physical states are given by 
maps from physical space into some non-linear space of field values, 
like, e.g., a sphere. A particular such model is the Skyrme model 
\cite{Skyrme:1971} in which the target space is the three-sphere 
$S^3$. Configurations of finite energy must map spatial infinity 
(physical space is $\reals^3$) into a single point of $S^3$ so 
that $Q$ decomposes into a countably infinity of path components 
according to the winding number of that map. In the Skyrme model, 
which serves to give an effective description of baryons, this 
winding number corresponds to the baryon number. The fundamental 
group of each path component is isomorphic to the fourth 
homotopy group of the target space $S^3$, which is again just 
$\integers_2$. One can now prove that the loops traced through by 
360-degree rotations are contractible in the components of even 
winding numbers and non-contractible in the components of odd winding 
numbers~\cite{Giulini:1993}. Hence spinorial states exist for odd
baryon numbers, as one should expect on physical grounds.   

These examples differ from those in General Relativity insofar as in 
the latter spinorial states usually exist only in non-abelian sectors, 
i.e. sectors that correspond to higher-dimensional unitary irreducible 
representations of the fundamental group~\cite{Giulini:1995a}. An
example will be mentioned below. 
For that reason we made the distinction between the first and the 
second (geometric quantisation) method of generalising Schr\"odinger 
quantisation, since non-abelian sectors are obtained in the first, 
but not in the second method, which is only sensitive to the 
abelianisation of the fundamental group. That the restriction to 
abelian sectors is unnecessary and unwarranted is further discussed 
in~\cite{Giulini:1995b} 

\begin{figure}[ht]
\centering
\includegraphics[width=0.9\linewidth]{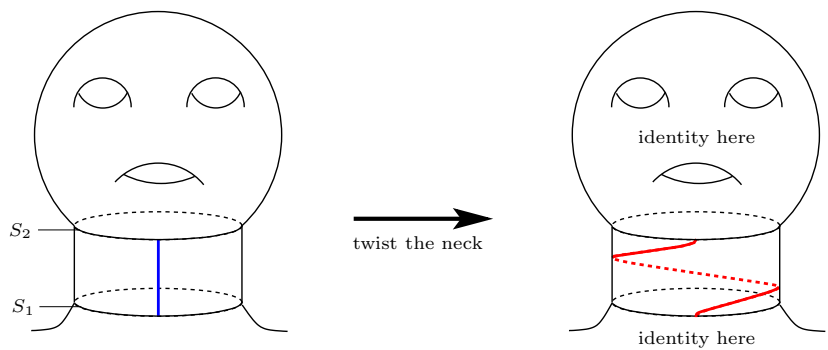}
\put(-72,-4){\scriptsize identity here}
\put(-72,72){\scriptsize identity here}
\put(-309,8){\scriptsize $S_1$}
\put(-310,37){\scriptsize $S_2$}
\put(-180,32){\scriptsize twist the neck}
\caption{\label{fig:FullRotation1}%
A full 360-degree rotation of the part of the manifold above 
the 2-sphere $S_2$ relative to the part below the 2-sphere $S_1$
is given by a diffeomorphism with support on the cylinder region
bounded by $S_1$ and $S_2$ that rotates one bounding sphere relative
to the other by 360 degrees (`twisting the neck' by 360 degrees).} 
\end{figure}

The geometric-topological situation underlying the existence of 
spinorial states in General Relativity is 
this~\cite{Friedman.Sorkin:1980}: Consider a 3-manifold $\Sigma$ 
with one regular end, so as to describe an asymptotically flat 
isolated system without internal infinities. Here `regular' means that that the one-point
compactification $\bar\Sigma$ of $\Sigma$ is again a manifold. 
This means that $\Sigma$ contains a compact subset the 
complement of which is a cylinder $\reals\times S^2$. A physical 
rotation of the system so represented is then given by a 
diffeomorphism whose support is entirely on that cylinder and 
rotates the $S^2$ at one end relative to the $S^2$ at the 
other end by full 360 degrees; see Fig.\,\ref{fig:FullRotation1}. 
\begin{figure}[ht]
\centering
\includegraphics[width=0.9\linewidth]{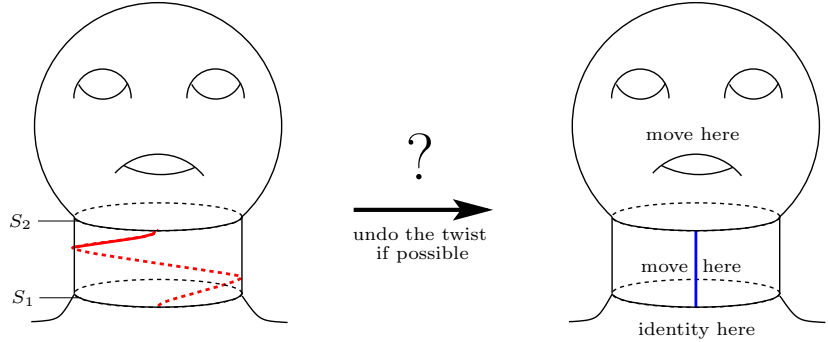}
\put(-72,-4){\scriptsize identity here}
\put(-69,70){\scriptsize move here}
\put(-71,20){\scriptsize move\ \  here}
\put(-309,8){\scriptsize $S_1$}
\put(-310,37){\scriptsize $S_2$}
\put(-160,55){\Huge ?}
\put(-180,32){\scriptsize undo the twist}
\put(-171,25){\scriptsize if possible}
\caption{\label{fig:FullRotation2}%
Keeping all points below $S_1$ fixed but now allowing the points 
above $S_2$ to move, the neck-twist may or may not be continuously 
undone through a continuous sequence of diffeomorphisms whose 
support is entirely above the sphere $S_1$. If it cannot be 
undone in this fashion, the manifold above $S_2$, or rather its 
one-point compactification, is called `spinorial'.
Do not be misled to think that you can just undo it by rigidly rotating 
the upper part in the embedding space shown here, since this 
will generally not define a diffeomorphism of the manifold
itself.} 
\end{figure}

The questions is now this: Is this diffeomorphism in the identity 
component of those diffeomorphisms that fix the 2-sphere at 
`spatial infinity'? See Fig.\,\ref{fig:FullRotation2} for further
illustration. 
The answer to this question just depends on 
the topology on $\Sigma$ and is now known for all 3 manifolds.%
\footnote{To decide this entails some subtle issues, like 
whether to diffeomorphisms that are homotopic (continuously 
connected through a one-parameter family of \emph{continuous}
maps) are also isotopic (continuously connected through 
a one-parameter family of \emph{homeomorphisms}) and then 
also diffeotopic (continuously connected through a one-parameter 
family of \emph{diffeomorphims}). The crucial question is 
whether homotopy implies isotopy, which is not at all obvious 
since on a homotopy the interpolating maps connecting two 
diffeomorphism are just required to be continuous, that
is, they need not be continuously invertible as for an 
isotopy. For example, the inversion $I(\vec x)=-\vec x$
in $\reals^n$ is clearly not isotopic to the identity, but 
homotopic to it via $\phi_t(\vec x)=(1-2t)\vec x$ for 
$t\in [0,1]$. Then $\phi_0=\mathrm{id}$, $\phi_1=I$, and 
only at $t=1/2$ does the map $\phi_t$ cease to be invertible.} 
Roughly speaking, the generic case is that spinorial spates 
are allowed. More precisely, those 3-manifolds $\bar\Sigma$ 
(from now on we represent the manifolds by their one-point 
compactifications in order to talk about closed spaces) which 
do not allow for spinorial states are connected sums of lens 
spaces and handles ($S^1\times S^2$). This is a very nice 
(though rather non-trivial) result insofar, as the non-spinoriality of these 
spaces as well as their connected sums is easy to visualise. 
Hence one may say that there are no other non-spinorial 
manifolds than the `obvious' ones.

Take, for example, the simplest lens space%
\footnote{\label{fnote:LensSpace}
The definition of lens spaces $L(p,q)$ in 3 dimensions is 
$L(p,q)=S^3/\!\!\sim$, where $(p,q)$ is a pair of positive 
coprime integers with $p>1$, 
$S^3=\{(z_1,z_2)\in\complex^2\mid\vert z_1\vert^2+\vert z_2\vert^2=1\}$, 
and $(z_1,z_2)\sim (z'_1,z'_2)\Leftrightarrow z'_1=\exp(2\pi i/p)z_1$, 
and $z'_2=\exp(2\pi i\,q/p)z_2$. One way to picture the space is 
to take a solid ball in $\reals^3$ and identify each points on the upper 
hemisphere with a points on the lower hemisphere after a rotation by 
$2\pi q/p$ about the vertical symmetry axis. In this way each set 
of $p$ equidistant points on the equator is identified to a single 
point. The fundamental group of $L(p,q)$ is $\integers_p$, i.e. 
independent of $q$, and the higher homotopy groups are those of its
universal cover, $S^3$. This does, however, not imply that 
$L(p,q)$ is homotopy equivalent, or even homeomorphic, to $L(p,q)$.
The precise relation will be stated below.}
$L(2,1)$, which is just real projective 3-space $\RP^3$. It can 
be imagined as
a solid ball $B$ in $\reals^3$ with antipodal points on 
the 2-sphere boundary identified. Think of an inner 
point, say the centre, of $B$ as the point at infinity, 
surround it by a small spherical shell whose inner boundary 
is the 2-sphere $S_1$ and outer boundary the 2-sphere 
$S_2$ (above we called it a `cylinder' since its topology 
is $\reals\times S^2$). Now perform a full 360-degree
rotation of $S_1$ against $S_2$ with support inside 
the shell. Can this diffeomorphism been undone through 
a continuous sequence of diffeomorphims that fix all points 
inside the inner sphere $S_1$? Clearly it can: Just rigidly 
rotate the outside to undo it. The crucial point is that 
this rigid rotation is compatible with the boundary 
identification and hence does indeed define a diffeomorphism 
of $\RP^3$. Essentially the same argument applies to all 
other `obvious cases'.

\begin{figure}[ht]
\centering
\includegraphics[width=0.5\linewidth]{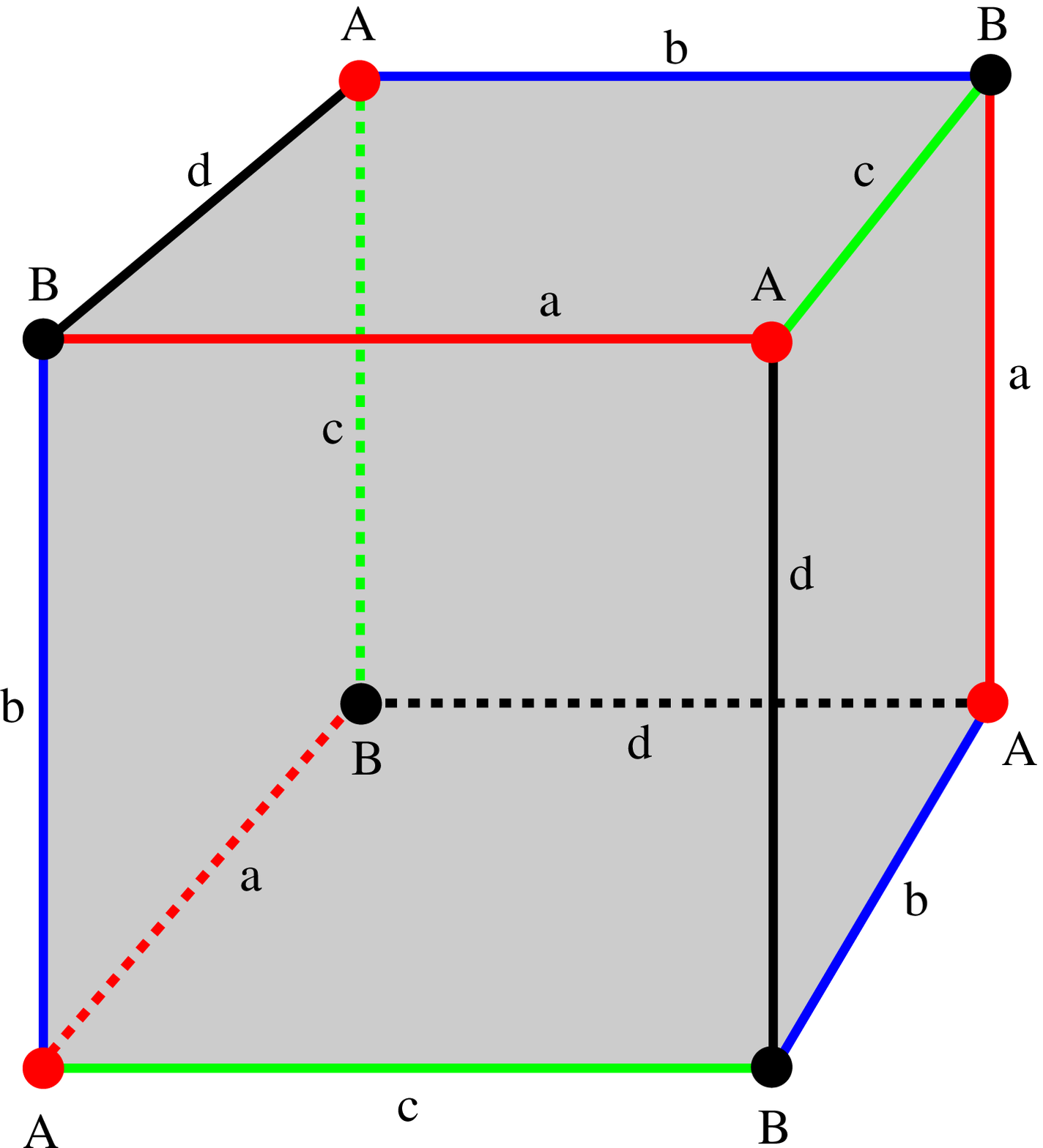}
\caption{\label{fig:Q-Space}%
The manifold $S^3/D_8^*$ is obtained from a solid cube
by identifying opposite faces after a relative 
90-degree rotation about the axis connecting their 
midpoints. In the picture shown here the identifying 
motion between opposite faces is a right screw, giving
rise to the identifications of edges and vertices 
as labelled in the picture.}   
\end{figure}

In contrast, it is much more difficult to prove that such an 
undoing is impossible, i.e. the spinoriality of a given manifold. 
Needless to say, the fact that you cannot easily visualise 
a possible undoing of a 360 degree twist does not mean it does
not exist. A simple and instructive example is given by the spherical 
space form $S^3/D_8^*$, where $D_8^*$ is the 8-element non-abelian 
subgroup of $SU(2)$ that doubly covers (via the double cover 
$SU(2)\rightarrow SO(3)$) the 4-element abelian subgroup 
of $SO(3)$ that is given by the identity and the three 
180-degree rotations about the mutually perpendicular 
$x$, $y$, and $z$ axes. Identifying $S^3$ with $SU(2)$ 
the quotient $S^3/D_8^*$ is defined by letting $D_8^*$
act through, say, right translations. Since $SU(2)$ is also 
the group of unit quaternions, $D_8^*$ can be identified 
with its subgroup $\{\pm 1,\pm i,\pm j,\pm k\}$, where 
$i,j,k$ denote the usual unit quaternions (they square 
to $-1$ and $ij=k$ and also cyclic permutations thereof).   
A way to visualise $S^3/D_8^*$ is given in 
Fig.\,\ref{fig:Q-Space}.    
Note that if the 2-dimensional boundary of the cube is 
smoothly deformed to a round 2-sphere a rigid rotation 
in the embedding $\reals^3$ would still not be 
compatible with the boundary identifications. In fact, 
it is known that $S^3/D_8^*$ is spinorial (see 
\cite{Giulini:1994a} for more information and 
references). Here we just remark that $D_8^*$ has five
equivalence classes of unitary irreducible representations:
Four one-dimensional and a single two-dimensional one.
The spinorial sector is that corresponding to the latter,
that is, it is a non-abelian sector.  

Another remarkable property of $S^3/D_8^*$ is that it is 
\emph{chiral}, that is, it does not admit for orientation 
reversing self-diffeomorphisms; see \cite{Giulini:1994a} for 
some information which 3-manifolds are chiral and 
\cite{Muellner:2008} for a recent systematic investigation 
of chirality in all dimensions. This means that if 
we had chosen the map that identifies opposite faces 
of the cube shown in Fig.\,\ref{fig:Q-Space} to be a left
rather than right screw, we would have obtained a manifold 
that is not orientation-preserving diffeomorphic to 
the the one originally obtained, though they are 
clearly orientation-reversing diffeomorphic as they 
are related by a simple reflection at the origin of the 
embedding $\reals^3$. Being chiral seems to be more the 
rule than the exception for 3-manifolds~\cite{Giulini:1994a}.

\section{Further developments}
\label{sec:FurtherDevelopments}
In the last subsection we have learnt that the fundamental 
group of the configuration space of the gravitational field 
will give rise to sectors with potentially interesting physical 
interpretations. Hence it seems natural to generally ask:
What is the fundamental group of the 
configuration space associated to a manifold $\Sigma$?
The last question can be given an elegant abstract answer,
though not one that will always allow an easy characterisation 
(determination of the isomorphicity class) of the group. 
The abstract answer is in terms of a presentation of a
certain mapping-class group and comes about as follows: Consider the 
3-manifold $\Sigma$ which we assume to have one regular 
end. Hence its one-point compactification, $\bar\Sigma$, 
is a manifold. Next consider the group of diffeomorphims 
$\DiffF$ that fix a prescribed point $p\in\bar\Sigma$
as well as all vectors in the tangent space at this 
point. It is useful to think of $p$ as the `point at 
infinity', i.e. the point that we added for 
compactification, for then it is intuitively clear that 
$\DiffbarF$ corresponds to those diffeomorphism of 
$\Sigma$ that tend to the identity as one moves to 
infinity within the single end. In order to have that
picture in mind, we will from now on write $\infty$ for 
the added point $p$. The configuration space
of the gravitational field on $\Sigma$ can then be 
identified with the space of Riemannian metrics on 
$\bar\Sigma$, $\Riembar$, modulo the identifications 
induced by $\DiffbarF$, i.e. 
\begin{equation}
\label{eq:ConfSpace}
Q(\Sigma)=\Riembar/\DiffbarF\,.
\end{equation}
Now, it is true that $\DiffbarF$ acts freely on $\Riembar$
(there are no non-trivial isometries on a Riemannian manifold 
that fix a point and the frame at that point) and that this 
action admits a slice (see \cite{Ebin:1968}). Hence $\Riembar$
is a principle fibre bundle with group $\DiffbarF$  and base 
$Q(\Sigma)$ (\cite{Fischer:1970,Fischer:1986}). But $\Riembar$ 
is contractible (being an open positive convex cone in the 
vector space of smooth sections of symmetric tensor fields 
of rank two over $\bar\Sigma$). Hence the long exact-sequence
of homotopy groups for the fibration 
$\DiffbarF\rightarrow\Riembar\rightarrow Q(\Sigma)$ implies
the isomorphicity of the $n\,$th homotopy group of the fibre
$\DiffbarF$ with the $n+1\,$st homotopy group of the base 
$Q(\Sigma)$ . In particular, the first homotopy group (i.e.
the fundamental group) of $Q(\Sigma)$ is isomorphic to the 
zeroth homotopy group of the group $\DiffbarF$. However, the latter 
is just the quotient $\DiffbarF/\DiffbarFconn$, where 
$\DiffbarFconn\subset\DiffbarF$ is the normal subgroup formed 
by the connected component of $\DiffbarF$ that contains the 
identity. In this way we finally arrive at the result that 
the fundamental group of $Q(\Sigma)$ is isomorphic to a 
\emph{mapping-class} group:
\begin{equation}
\label{eq:HomotopyIsom}
\pi_1\bigl(Q(\Sigma)\bigr)\cong \DiffbarF/\DiffbarFconn\,.
\end{equation}

This is a very interesting result in its own right. It contains the 
mathematical challenge to characterise $\DiffbarF/\DiffbarFconn$. 
A way to attack this problem is to use the fact that each element 
in $\DiffbarF$ defines a self-map $\pi_1(\bar\Sigma,\infty)$ just 
by mapping loops based at $\infty$ to their image loops, which are
again based at $\infty$ since elements of $\DiffbarF$ keep that
point fixed. Since in this fashion homotopic loops are mapped to 
homotopic loops, this defines indeed a map on 
$\pi_1(\bar\Sigma,\infty)$ which is, in fact, an automorphism. 
Moreover, elements in the identity component $\DiffbarFconn$ 
give rise to the trivial automorphisms. This is obvious, since 
the images of a loop under continuously related diffeomorphisms
will in particular result in homotopic loops. Hence we have in 
fact a homomorphism from $\DiffbarF/\DiffbarFconn$ into the 
automorphism group of $\pi_1(\bar\Sigma,\infty)$:
\begin{equation}
\label{eq:HomoIntoAut}
h:\DiffbarF/\DiffbarFconn\rightarrow
\mathrm{Aut}\bigl(\pi_1(\bar\Sigma,\infty)\big)\,.
\end{equation}
The strategy is now this: Assume we know a presentation of 
$\mathrm{Aut}\bigl(\pi_1(\bar\Sigma,\infty)\big)$, that is, a 
characterisation of this group in terms of generators and relations. 
Then we aim to make useful statements about the kernel and image 
of the map in (\ref{eq:HomoIntoAut}) so as to be able to 
derive a presentation for $\DiffbarF/\DiffbarFconn$. A simple but 
non-trivial example will be given below. We recall that $\bar\Sigma$ is a 
unique connected sum of prime manifolds and that $\pi_1(\bar\Sigma)$
is the free product of the fundamental groups of the primes. Since
(finite) presentations for the automorphism group of a free product
can be derived if (finite) presentations for the automorphism groups
of the factors are known~\cite{Gilbert:1987}, we in principle only 
need to know the latter.        

Another mathematically interesting aspect connected with 
(\ref{eq:HomotopyIsom}) is the fact that $\DiffbarF/\DiffbarFconn$ 
is a topological invariant of $\bar\Sigma$ which is \emph{not} a 
homotopy invariant~\cite{McCarty:1963}. Hence (\ref{eq:HomotopyIsom})
implies that $\pi_1\bigl(Q(\Sigma)\bigr)$, too, is a topological 
invariant of $\bar\Sigma$ which is not homotopy invariant, i.e. it might 
tell apart 3-manifolds which are homotopically equivalent but not 
homeomorphic. There are indeed examples for this to happen. Here is 
one: Recall that lens spaces (see footnote~\ref{fnote:LensSpace})
$L(p,q)$ and $L(p,q')$ are homotopy equivalent iff 
$qq'=\pm n^2\,(\mathrm{mod}\,p)$ for some integer $n$ 
(theorem\,10 in \cite{Whitehead:1941}) and homeomorphic%
\footnote{As regards the notion of chirality, an interesting 
refinement of this statement is that $L(p,q)$ and $L(p,q')$ are 
\emph{orientation-preserving} homeomorphic iff 
$q'=q^{\pm 1}\,(\mathrm{mod}\,p)$~\cite{Reidemeister:1935}.} 
iff (all four possibilities) $q'=\pm q^{\pm 1}\,(\mathrm{mod}\,p)$
(here all four possibilities of combinations of $\pm$ signs are 
considered). On the other hand, the mapping-class group 
$\DiffbarF/\DiffbarFconn$ for $L(p,q)$ is $\integers\times\integers$ 
if $q^2=1\,(\mathrm{mod}\,p)$ with $q\not =\pm 1\,(\mathrm{mod}\,p)$ 
and $\integers$ in the remaining cases for $p>2$ 
(see Table\,IV on p.\,591 of~\cite{Witt:1986b}). Take now, as an 
example, $p=15$, $q=1$, and $q'=4$. Then the foregoing implies 
that $L(15,1)$ and $L(15,4)$ are homotopic but not homeomorphic 
and have different mapping-class groups.   

\begin{figure}[ht]
\centering
\includegraphics[width=0.55\linewidth]{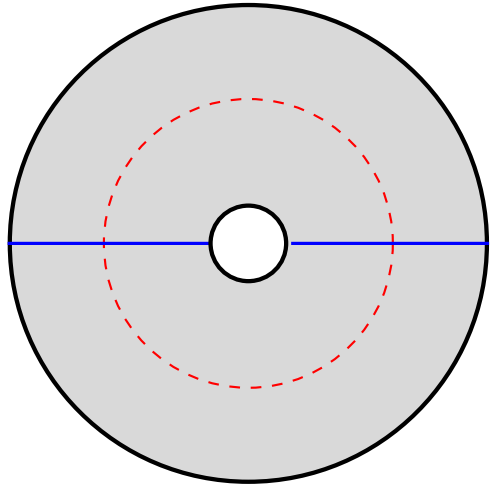}
\caption{\label{fig:RP3x2}
The connected sum $\RP^3\#\RP^3$ between two real projective spaces 
may be visualised as a spherical shell (here the grey-shaded region) 
where antipodal points on each of the two 2-sphere boundaries, $S_1$ 
and $S_2$, are identified. The 2-dimensional figure here should be 
rotated about the horizontal symmetry axis. The two horizontal line 
segments shown form a circle in view of the antipodal identifications. 
It shows that $\RP^3\#\RP^3$ is a circle bundle over $\RP^2$. The dotted 
circle, which upon rotation of the figure becomes a 2-sphere can be 
thought of as the 2-sphere along which the connected sum between the 
two individual $\RP^3$ manifolds is taken.}
\end{figure}

Finally we give an example for a presentation and its pseudo-physical 
interpretation for $\DiffbarF/\DiffbarFconn$. Consider the connected 
sum (denoted by $\#$) of two real projective spaces $\RP^3$. 
This manifold may be visualised as explained in Fig.\,\ref{fig:RP3x2}. 

\begin{figure}
\begin{minipage}[b]{0.45\linewidth}
\centering\includegraphics[width=0.9\linewidth]{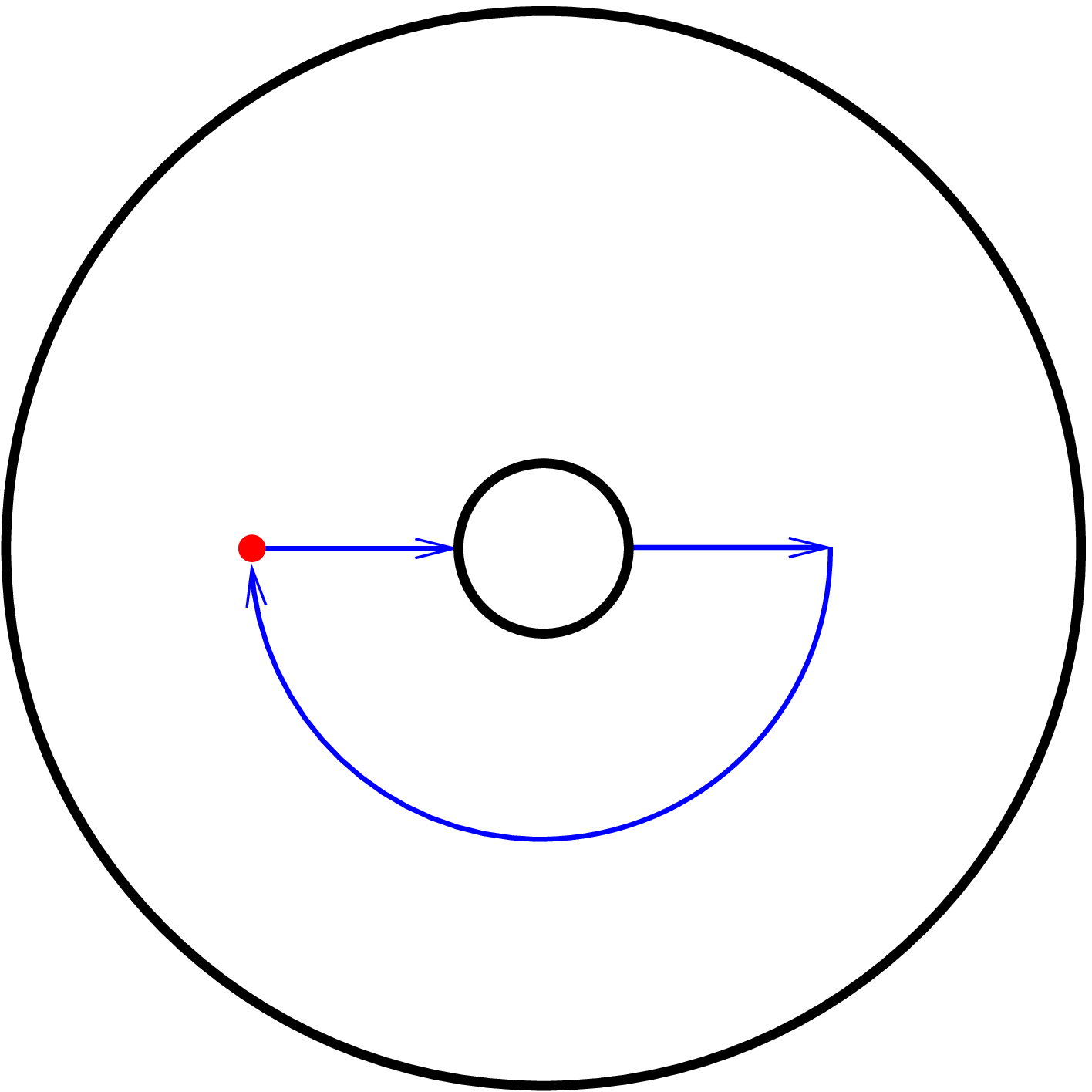}
\put(-75,95){\huge $a$}
\put(-114,67){$p$}
\end{minipage}
\hfill
\begin{minipage}[b]{0.45\linewidth}
\centering\includegraphics[width=0.9\linewidth]{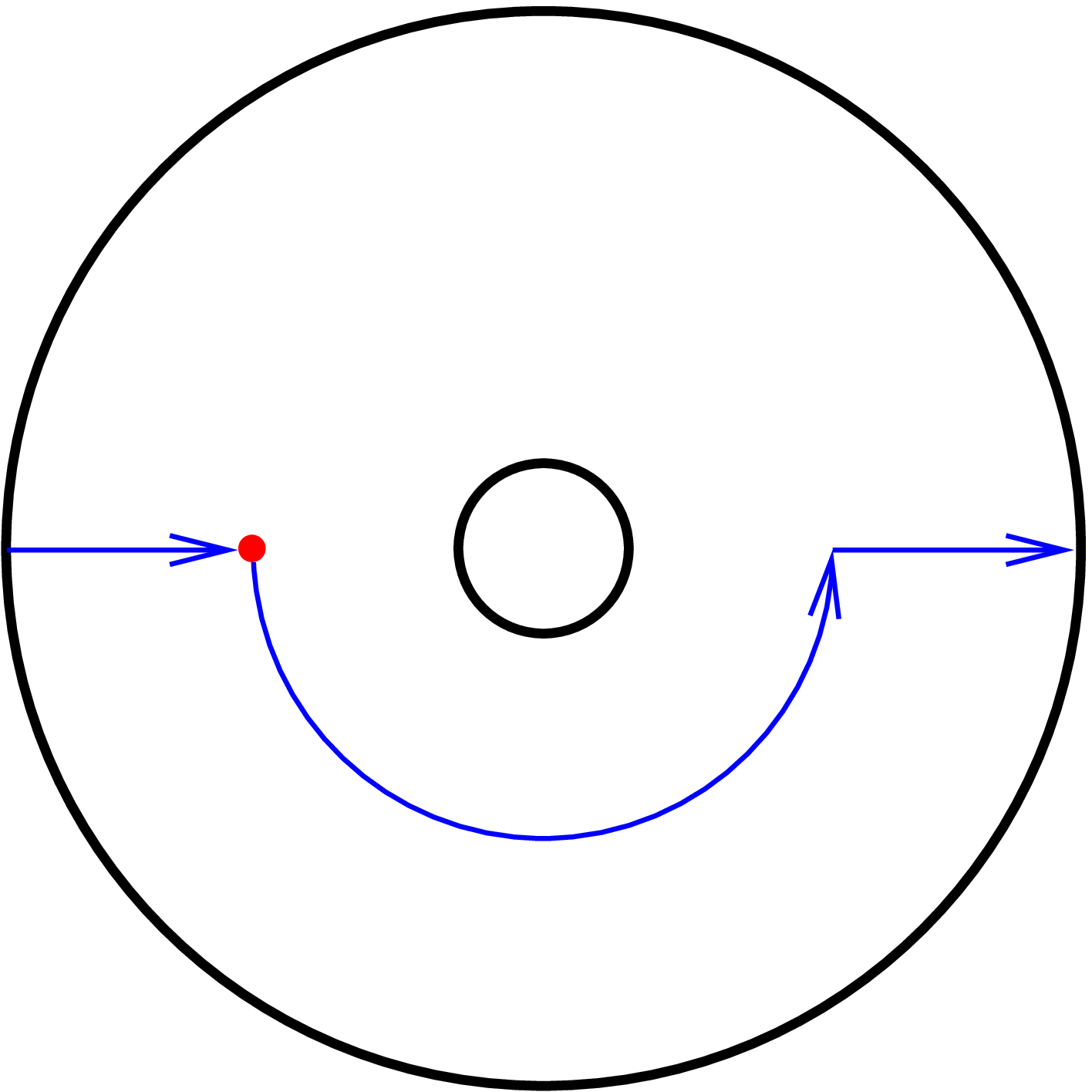}
\put(-73,95){\huge $b$}
\put(-100,67){$p$}
\end{minipage}
\caption{\label{fig:FundGroupGen}%
Shown are two closed loops based at some point $p$ 
whose homotopy classes generate the fundamental group 
$\integers_2 *\integers_2\cong \integers_2\ltimes\integers$ 
of the manifold $\RP^3\#\RP^3$.} 
\end{figure}
The fundamental group of $\RP^3\#\RP^3$ is the twofold free 
product $\integers_2 *\integers_2$ of the fundamental group 
$\integers_2$ of the factors $\RP^3$. With respect to the 
generators $a,b$ shown in Fig.\,\ref{fig:FundGroupGen} or, 
alternatively, with respect to the generators $a,c$, where $c:=ab$
corresponds to the loop shown by the two horizontal segments in 
Fig.\,\ref{fig:RP3x2}, two alternative presentation of the 
fundamental group are given by  
\begin{equation}
\label{eq:FundGroup}
\pi_1(\RP^3\#\RP^3)=
\underbrace{\langle a,b\mid a^2=b^2=1\rangle}_%
{\cong\,\integers_2\times\integers_2}=
\underbrace{\langle a,c\mid a^2=1,\ aca^{-1}=c^{-1}\rangle}_%
{\cong\,\integers_2\ltimes\integers}
\end{equation}

Turning to (\ref{eq:HomoIntoAut}) we first remark that the 
automorphism group of $\integers_2*\integers_2$ is itself 
isomorphic to $\integers_2*\integers_2$,
\begin{equation}
\label{eq:AutTwoZtwos}
\mathrm{Aut}(\integers_2*\integers_2)\cong\integers_2*\integers_2
=\langle E,S\mid E^2=S^2=1\rangle\,,
\end{equation}
where the two generators $E,S$ can be identified by stating 
their action on the generators $a,b$ of the fundamental 
group:
\begin{equation}
\label{eq:AutTwoZtwosGen}
E: (a,b)\mapsto (b,a)\,,\quad
S: (a,b)\mapsto (a,aba^{-1})\,.
\end{equation}
It may now be shown that the map $h$ in (\ref{eq:HomoIntoAut}) is 
an isomorphism so that the fundamental group of the configuration 
space $Q(\Sigma)$ is the free product $\integers_2 *\integers_2$. 
Injectivity of $h$ is not so obvious (but true) whereas surjectivity 
can be shown by visualising diffeomorphisms that actually realise 
the generators $E$ and $S$ of (\ref{eq:AutTwoZtwosGen}). 
For example, $E$ can be realised by an inversion on the sphere 
along which the connected sum is taken (see Fig\,\ref{fig:RP3x2}) 
(which is orientation reversing) followed by a simple reflection 
along a symmetry plane (so as to restore orientation preservation). 
Its `physical' meaning is that of an exchange of the two diffeomorphic
factors (primes) in the connected-sum decomposition. The map for 
$S$ is a little harder to visualise since it mixes points between 
the two factors; see \cite{Giulini:2007a} for pictures. 
It can roughly be described as sliding one factor through the 
other and back to its original position. Here we wish to focus 
on the following: Given the generalisations of Schr\"odinger
quantisations outlined above, we are naturally interested 
in the equivalence classes of unitary irreducible representations 
of $\integers_2 *\integers_2$. They can be obtained by elementary 
means and are represented by the four obvious one-dimensional 
representations where $E\mapsto\pm 1$ and $S\mapsto\pm 1$, and 
a continuum of 2-dimensional ones where 
\begin{equation}
\label{eq:MCGRepTwoRP3s}
E\mapsto
\left(
\begin{array}{lr}
1&0\\ 0&-1
\end{array}\right)\,,\qquad
S\mapsto
\left(
\begin{array}{lr}
\cos\tau & \sin\tau\\ 
\sin\tau & -\cos\tau\\
\end{array}\right)\,,\quad
\tau\in(0,\pi)\,.
\end{equation}
No higher dimensional ones occur.
The one-dimensional representations already show that both 
statistics sectors exist. Moreover, the two-dimensional 
representations show that the diffeomorphims representing $S$
mix the statistics sectors by an angle $\tau$ that depends on 
the representation class. All this may be read as indication 
against a classical `spin-statistics correlation' that one might have 
expected from experience with other non-linear field theories,
e.g. following \cite{Finkelstein.Rubinstein:1968}\cite{Sorkin:1988}.
Such a connection can therefore only exist in certain sectors
and the question can (and has) be asked how these sectors are 
selected~\cite{Dowker.Sorkin:1998,Dowker.Sorkin:2000}. 
See \cite{Giulini:1994a,Giulini:1997a} for other examples
with explicit presentations of $\DiffbarF/\DiffbarFconn$
where $\bar\Sigma$ is either the $n$ fold connected sum of 
real projective spaces $\RP^3$ or handles $S^1\times S^2$ and 
also some general statements. 

From what has been said so far it clearly emerges that the enormous 
topological variety and complexity of 3-manifolds leave their structural 
traces in General Relativity, which can be used to model some of the 
properties in pure gravity that are usually associated with ordinary
matter. This is indeed made practical use of, e.g. in modelling 
scattering and merging processes of black holes with data corresponding 
to wormhole topologies. But one should also say that the physical 
relevance of much of what I said later is not at all established. 
The aim of my presentation was to alert to the existence of these 
structures, leaving their physical relevance open for the time being. 
Somehow all this may remind one Tait's beautiful idea to model the 
discrete structural properties of material atoms on the properties 
of knots in physical space, which he thought of as knotted vortex 
lines in the all-embracing hypothetical ether medium. But whereas 
there was never formulated a fundamental dynamical theory of the 
ether\footnote{%
Maxwell's equations were thought of as a kind of effective theory 
that describes things on a coarse-grained scale, so that e.g. the 
vortex knots could be approximated by point particles.} there is 
a well formulated and well tested theory of geometrodynamics: General 
Relativity. In that sense we are in a much better position than 
Tait was in the mid 1880s.

\vspace{1.0 cm}
\noindent
\textbf{Acknowledgements}
I sincerely thank the organisers of the \emph{Beyond-Einstein} 
conference at Mainz University for inviting me to this most
stimulating and pleasant meeting.

\bibliographystyle{plain}
\bibliography{RELATIVITY,HIST-PHIL-SCI,MATH,QM} 
\end{document}